\documentstyle[jkas]{article}

\beginpage{1}
\endpage{25}
\year{2013}\volume{00}\month{December}

\runningauthor {S. TRIPPE} 
\runningtitle{POLARIZATION AND POLARIMETRY} 

\month{December} \year{2013} \volume{00} \issueno{00}
\beginpage{1}\endpage{25}
\date{Received 30 August 2013; Revised 17 December 2013; Accepted 28 December 2013}

\begin{document}


\def\E{{\bf E}}
\def\B{{\bf B}}
\def\e{{\bf e}}
\def\J{{\bf J}}
\def\SS{{\bf S}}
\def\M{{\bf M}}
\def\T{{\bf T}}
\def\C{{\bf C}}
\def\al{{\langle}}
\def\ar{{\rangle}}
\def\deg{{^{\circ}}}
\newcommand\name[1]{{\small\sc #1}}
\newcommand\para[1]{{\noindent{\bf #1}}}
\newcommand\epara[1]{{\noindent{\it #1}}}
\newcommand\myion[2]{{#1\,{\sc #2}}}

\makeatletter
\renewcommand*\l@section{\@dottedtocline{1}{0em}{2em}}
\renewcommand*\l@subsection{\@dottedtocline{2}{2em}{3em}}
\renewcommand*\l@subsubsection{\@dottedtocline{3}{5em}{4em}}
\makeatother

\let \savenumberline \numberline
\def \numberline#1{\savenumberline{#1.}}


\title{POLARIZATION AND POLARIMETRY: A REVIEW}

\author{Sascha Trippe}

\address{Department of Physics and Astronomy, Seoul National University, Seoul 151-742, South Korea\\ {\it E-mail: trippe@astro.snu.ac.kr}}

\address{\normalsize{\it (Received 30 August 2013; Revised 17 December 2013; Accepted 28 December 2013)}}

\abstract{\noindent Polarization is a basic property of light and is fundamentally linked to the internal geometry of a source of radiation. Polarimetry complements photometric, spectroscopic, and imaging analyses of sources of radiation and has made possible multiple astrophysical discoveries. In this article I review (i) the physical basics of polarization: electromagnetic waves, photons, and parameterizations; (ii) astrophysical sources of polarization: scattering, synchrotron radiation, active media, and the Zeeman, Goldreich-Kylafis, and Hanle effects, as well as interactions between polarization and matter (like birefringence, Faraday rotation, or the Chandrasekhar-Fermi effect); (iii) observational methodology: on-sky geometry, influence of atmosphere and instrumental polarization, polarization statistics, and observational techniques for radio, optical, and X/$\gamma$ wavelengths; and (iv) science cases for astronomical polarimetry: solar and stellar physics, planetary system bodies, interstellar matter, astrobiology, astronomical masers, pulsars, galactic magnetic fields, gamma-ray bursts, active galactic nuclei, and cosmic microwave background radiation.}


\maketitle

\small
\tableofcontents
\normalsize

\section{INTRODUCTION \label{sect_intro}}  

\begin{quote}
\footnotesize
{\bf ``}\,To many astrophysicists, stellar polarimetry is a Cinderella subject considered as being so insignificant and, at the same time, being so esoteric as to be ignored and left alone. [...] There can be no doubt, however, that the study of polarization within astronomy has a strong role to play either in its own right, or in combination with other observational tools, as a diagnostic for understanding the behaviour of celestial sources.\,{\bf ''}

--- \citet{clarke2010}, p.\,XIII
\end{quote}

\noindent
Historically, the study of polarized light began with the discovery of birefringence in crystals by Erasmus Bartholinus and its subsequent interpretation by Christian Huygens around the year 1670 \citep{brosseau1998}. Astronomical observations of polarized light commenced in the middle of the 19th century; some of the earliest publications treat the linear polarization of sun light reflected by the moon \citep{secchi1860} and the linear polarization of the light from the solar corona \citep{edlund1860}. Subsequently, the field of polarimetry evolved closely with the technical progress of observational techniques in general: from optical polarimetry to radio polarimetry in the 1940s \citep{wilson2010} and eventually to space-based X-ray polarimetry in the 1970s \citep{weisskopf1978}.

Polarization is a fundamental property of electromagnetic radiation. It is a rich source of information on the physical properties -- magnetic fields, internal conditions, particle densities, et cetera -- of astronomical objects. Polarimetric observations complement analysis methods based on photometry as well as on spectral (spectroscopy) or angular (mapping, imaging) resolution. Accordingly, polarimetry has contributed substantially to the progress of astronomy. Milestones have been, for example:

\begin{itemize}

\item  the mapping of solar and stellar magnetic fields (e.g., \citealt{schrijver2000});

\item  the characterization of the surface composition of solar system bodies (e.g., \citealt{bowell1974});

\item  the discovery of synchrotron radiation from astronomical objects \citep{oort1956};

\item  the discovery and characterization of large-scale (many kpc) galactic magnetic fields (e.g., \citealt{kulsrud2008});

\item  the analysis of the polarization modes of the cosmic microwave background (e.g., \citealt{kovac2002}).

\end{itemize}

This review provides a broad overview on the theory and phenomenology of polarization in astronomy. It covers (i) the physical basics of polarized radiation, (ii) sources of astrophysical polarization, (iii) concepts and methods of astronomical polarimetry, and (iv) astrophysical science cases.

\section{PHYSICAL BASICS \label{sect_basics}}  

\subsection{Electromagnetic Waves \label{ssect_waves}}

\subsubsection{Electric Field Vectors \label{sssect_efields}}

\noindent
The concept of electromagnetic waves derives from non-zero solutions of Maxwell's equations in vacuum, meaning here specifically the absence of electric charges (e.g., \citealt{landau1997,jackson1999}). Following common conventions, we regard the electric field\footnote{The magnetic field is perpendicular to the direction of travel and to the electric field. The amplitude of the magnetic field, $B$, is related to the amplitude of the electric field, $E$, like $B=E/c$.} $\bf E$ of an electromagnetic wave traveling in $z$ direction in Euclidean coordinates $(x,y,z)$ with speed of light $c$. Accordingly, we have -- in trigonometric notation --

\begin{equation}
\E(t,z) = \E(0,0)\cos(\omega t - kz - \phi)
\label{eq_wave}
\end{equation}

\noindent
where $t$ is the time, $\omega$ denotes the angular frequency, $k=\omega/c$ is the absolute value of the wave vector, and $\phi$ denotes an arbitrary phase. As the electric field vector is perpendicular to $z$, we can decompose $\E(t,z)$ into its $x$ and $y$ components. For simplicity, we regard the location $z=0$ only, i.e., we regard the location of $\E(t)$ in the $xy$ plane. The $x$ and $y$ components are then

\begin{eqnarray}
\label{eq_xywaves}
E_x(t) &=& E_x(0)\cos(\omega t - \phi_1)     \\
E_y(t) &=& E_y(0)\cos(\omega t - \phi_2) ~ . \nonumber
\end{eqnarray}

\noindent
Here, $\phi_{1,2}$ denote two -- a priori arbitrary -- phases. In addition, we denote the angle between $\E(t)$ and the positive $x$ axis -- the \emph{polarization angle}, counted in counterclockwise direction -- with $\chi$. The \emph{polarization} of the wave is given by the relative values of $E_x(0), E_y(0), \phi_1$, and $\phi_2$ \citep{rybicki1979,huard1997,landau1997,jackson1999,born1999,goldstein2003}.

\subsubsection{Elliptical Polarization \label{sssect_ellpol}}

\noindent
In general, the tip of the electric field vector follows an elliptical trajectory in the $xy$ plane; accordingly, the light is denoted as \emph{elliptically polarized}. The orientation of the ellipse in the $xy$ plane is constant in time; the polarization angle $\chi$ corresponds to the angle between the positive $x$ axis and the semi-major axis of the ellipse (counted in counterclockwise direction).

Elliptical polarization is the most general state of polarization of an electromagnetic wave. \emph{Linear} polarization occurs if the polarization ellipse degenerates into a line. \emph{Circular} polarization corresponds to the -- opposite -- special case of the ellipse degenerating into a circle.

\subsubsection{Linear Polarization \label{sssect_linpol}}

\noindent
For the case $\phi_1=\phi_2$, using here specifically $\phi_1=\phi_2=0$ without loss of generality, we have

\begin{eqnarray}
\label{eq_linpolwaves}
E_x(t) &=& E_x(0)\cos(\omega t)     \\
E_y(t) &=& E_y(0)\cos(\omega t) ~ . \nonumber
\end{eqnarray}

\noindent
The orientation of $\E$ then depends only on the magnitudes of $E_x(0)$ and $E_y(0)$ and is independent of time; the angle $\chi$ is constant. The radiation is \emph{linearly polarized} with polarization angle $\chi\in[0,\pi]$. The orientation of the plane wherein the wave is located -- the \emph{plane of polarization} -- as given by $\chi$ has an orientation but no direction; accordingly, the location of linear polarization in the $xy$ plane is not a vector. In physics and astronomy, the $x$ and $y$ components of linearly polarized light are commonly identified with \emph{horizontal} (H) and \emph{vertical} (V) polarizations, respectively.

\subsubsection{Circular Polarization \label{sssect_circpol}}

\noindent
In case of a relative phase shift $\phi_2=\phi_1\pm\pi/2$, using here specifically $\phi_1=0$ without loss of generality, \emph{and} $E_x(0)=E_y(0)$, we have

\begin{eqnarray}
\label{eq_circpolwaves}
E_x(t) &=& E_x(0)\cos(\omega t)     \\
E_y(t) &=& \pm E_y(0)\sin(\omega t) ~ . \nonumber
\end{eqnarray}

\noindent
The tip of the electric field vector moves circularly in the $xy$ plane with angular frequency $\omega$: the radiation is \emph{circularly polarized}. The sign of $E_y(t)$ -- which derives from the relative phase -- determines the sense of the motion of $\E$. A \emph{positive} sign, corresponding to counterclockwise motion, is commonly referred to as \emph{right-hand circular} (RHC) polarization. Accordingly, a \emph{negative} sign, corresponding to clockwise motion, is denoted as \emph{left-hand circular} (LHC) polarization.

\subsubsection{Macroscopic Polarization \label{sssect_macropol}}

\noindent
An individual electromagnetic wave is necessarily polarized as described above \emph{(microscopic polarization)}. Astrophysical observations do not deal with individual waves but with radiation that is a (almost always incoherent) superposition of a very large number of elementary electromagnetic waves. Accordingly, astronomical observations are sensitive to the \emph{macroscopic} polarization of light. For most physical systems, all orientations of the electric field vectors from the elementary emitters are equally probable -- there is, a priori, no reason to expect a macroscopic polarization of light. Confusingly, unpolarized light is therefore sometimes referred to as ``natural'' light.

A (macroscopic) polarization signal \citep{fowles1975,rybicki1979,mandel1995,born1999} can occur whenever the internal geometry of the source of radiation or the properties of the interstellar medium prefer a certain orientation of electric field vectors. In those cases -- discussed in detail in \S\,\ref{sect_polsources} -- the light becomes partially polarized with a \emph{degree of polarization}

\begin{equation}
m_P = \frac{I_P}{I} \in [0,1]
\label{eq_mP}
\end{equation}

\noindent
where $I$ denotes the \emph{total} intensity of the light and $I_P$ denotes the intensity of \emph{polarized} light.\footnote{In the astronomical literature, degrees of polarization are commonly quoted in units of per cent (\%).} Intensity $I$ and amplitude of electric field $E$ are related like $I\propto E^2$. The total polarization state of radiation can be described as a superposition of linear and circular polarization (cf. \S\,\ref{sssect_ellpol}--\ref{sssect_circpol}); accordingly, we can define separate degrees of linear and circular polarization. The degree of \emph{linear} polarization is given by

\begin{equation}
m_L = \frac{I_L}{I} \in [0,1]
\label{eq_mL}
\end{equation}

\noindent
where $I_L$ denotes the intensity of linearly polarized light. The degree of \emph{circular} polarization is given by

\begin{equation}
m_C = \frac{I_C}{I} \in [-1,1]
\label{eq_mC}
\end{equation}

\noindent
where $I_C$ denotes the intensity of circularly polarized light. The sign of $I_C$ and thus $m_C$ depends on the orientation of the polarization. By convention, the positive (negative) sign is assigned to light with $I_{\rm RHC}-I_{\rm LHC}>0$ $(<0)$; here, $I_{\rm RHC}$ and $I_{\rm LHC}$ denote the intensities of right-hand circularly and left-hand circularly polarized light, respectively \citep{hamaker1996b}.

\subsection{Photons \label{ssect_photons}}

\noindent
The wave-particle dualism of light \citep{einstein1905} implies that polarization is a property of individual photons; each photon can be assigned an individual state of polarization \citep{dirac1958}.

Using the standard \emph{bra--ket} notation for quantum states,\footnote{One may picture bras $\langle X|$ as complex row vectors and kets $|X\rangle$ as complex column vectors of equal -- potentially infinite -- dimension. A bra transforms into a ket by transposition plus complex conjugation (Hermitian conjugate). Accordingly, the product $|X\rangle\langle Y|$ corresponds to a complex matrix, the product $\langle X|Y\rangle$ corresponds to a complex scalar (e.g. \citealt{dirac1958}).} we may write two arbitrary photon states like $|X\rangle$ and $|Y\rangle$. These two states are \emph{orthogonal} if $\langle X|Y\rangle=0$. We may further assume that all states are \emph{normalized}, meaning $\langle X|X\rangle=1$ for arbitrary $X$. The bracket product $\langle X|Y\rangle$ is the \emph{probability amplitude} of the event ``The system in state $X$ is also in state $Y$'', and $|\langle X|Y\rangle|^2\in[0,1]$ is the corresponding \emph{probability}; $|...|$ denotes the absolute value of the enclosed function.

For the specific case of photon polarization \citep{bachor2004}, we have to consider photon states corresponding to horizontal linear polarization $|H\rangle$, vertical linear polarization $|V\rangle$, right-hand circular polarization $|R\rangle$, and left-hand circular polarization $|L\rangle$.\footnote{Only up to this sub-section $V$ denotes vertical linear polarization. In the remainder of this paper, $V$ denotes the corresponding Stokes parameter (\S\,\ref{sssect_stokes}).} Evidently, $|H\rangle$ and $|V\rangle$ on the one hand and $|R\rangle$ and $|L\rangle$ on the other hand are mutually exclusive, meaning

\begin{eqnarray}
\label{eq_photon-ortho}
\langle H|V\rangle = \langle V|H\rangle &=& 0     \\
\langle R|L\rangle = \langle L|R\rangle &=& 0 ~ . \nonumber
\end{eqnarray}

\noindent
Due to normalization, we further have

\begin{eqnarray}
\label{eq_photon-norm}
\langle H|H\rangle = \langle V|V\rangle &=& 1     \\
\langle R|R\rangle = \langle L|L\rangle &=& 1 ~ . \nonumber
\end{eqnarray}

\noindent
In addition, we have to note the relations between linear and circular polarization states. A standard tool employed in quantum optics is a polarizing beam-splitter that is sensitive to linear polarization: in a (thought) laboratory experiment, incident photons in state $|H\rangle$ are sent into one direction, photons in state $|V\rangle$ into another. The beam-splitter is insensitive to circular polarization; an incident photon in state $|R\rangle$ or $|L\rangle$ is sent into either direction with equal probability of 50\%. Accordingly, one finds

\begin{eqnarray}
\label{eq_photon-lin-vs-circ}
|\langle R|V\rangle|^2 = |\langle R|H\rangle|^2 &=& 0.5     \\
|\langle L|V\rangle|^2 = |\langle L|H\rangle|^2 &=& 0.5 ~ . \nonumber
\end{eqnarray}

\noindent
Combining the information from Eqs. \ref{eq_photon-ortho}, \ref{eq_photon-norm}, and \ref{eq_photon-lin-vs-circ}, it turns out \citep{bachor2004} that circular polarization states can be expressed as superpositions of linear polarization states like

\begin{eqnarray}
\label{eq_photon-circ-from-lin}
|R\rangle &=& \frac{1}{\sqrt{2}}\left(|H\rangle + i|V\rangle\right)     \\
|L\rangle &=& \frac{1}{\sqrt{2}}\left(|H\rangle - i|V\rangle\right) ~ . \nonumber
\end{eqnarray}

\noindent
with $i$ being the imaginary unit. An \emph{arbitrary} polarization state $|P\rangle$ can be expressed as

\begin{equation}
|P\rangle = a\,|H\rangle + e^{i\phi}\,b\,|V\rangle
\label{eq_photon-totalpol}
\end{equation}

\noindent
with $\phi$, $a$, and $b$ being real numbers, and $a^2+b^2=1$. The representation given by Eq. \ref{eq_photon-totalpol} is not unique; any pair of orthogonal states can be used as base vectors.

In analogy to the case of electromagnetic waves, the macroscopic polarization of light is given by the superposition of a large number of photons with individual microscopic polarization states. The discussion provided in \S\,\ref{sssect_macropol} is equally valid for waves and particles.

\subsection{Parameterizations \label{ssect_param}}

\subsubsection{Jones Calculus \label{sssect_jones}}

\noindent
In Eq. \ref{eq_xywaves}, I introduced the components of the electric field in trigonometric notation for convenience. Likewise, the electric field can be given in complex exponential notation as

\begin{eqnarray}
\label{eq_xyfield}
E_x(t) &=& E_x(0) e^{i(\omega t - \phi_1)}     \\
E_y(t) &=& E_y(0) e^{i(\omega t - \phi_2)} ~ . \nonumber
\end{eqnarray}

\noindent
One may now define \citep{jones1941,fowles1975,huard1997,goldstein2003} the \emph{Jones vector}

\begin{equation}
\label{eq_jones-vector}
\e \equiv \left[ \begin{array}{r} E_x(0)\,e^{i\phi_1} \\ E_y(0)\,e^{i\phi_2} \end{array} \right]
\end{equation}

\noindent
that expresses amplitudes and phases of the electric field in vector form. A convenient -- usually not normalized -- form of the Jones vector is achieved by expressing the components in units of the amplitude of one of them. Linearly polarized waves may be expressed like 
\begin{equation}
\label{eq_jones-lin}
\e_x = \left[\begin{array}{r} 1 \\ 0 \end{array}\right]; ~~ \e_y = \left[\begin{array}{r} 0 \\ 1 \end{array}\right]
\end{equation}

\noindent
where $\e_x$ and $\e_y$ denote waves polarized in $x$ and $y$ direction, respectively. Likewise, by exploiting the identity $\pm i=e^{\pm i\pi/2}$, circularly polarized waves can be expressed like

\begin{equation}
\label{eq_jones-circ}
\e_L = \left[\begin{array}{r} 1 \\ i \end{array}\right]; ~~ \e_R = \left[\begin{array}{r} 1 \\ -i \end{array}\right]
\end{equation}

\noindent
where $\e_L$ and $\e_R$ denote left-hand and right-hand circular polarization, respectively. The result of a \emph{superposition} of electric fields is given by the sum of the appropriate Jones vectors. A noteworthy example is

\begin{equation}
\label{eq_circ-lin}
\left[\begin{array}{r} 1 \\ i \end{array}\right] + \left[\begin{array}{r} 1 \\ -i \end{array}\right] = \left[\begin{array}{r} 2 \\ 0 \end{array}\right] = 2\left[\begin{array}{r} 1 \\ 0 \end{array}\right]
\end{equation}

\noindent
which demonstrates that a linearly polarized wave can be expressed as the sum of a left-hand and a right-hand circularly polarized wave with equal amplitudes -- as demanded by equivalence with Eqs. \ref{eq_photon-circ-from-lin}, \ref{eq_photon-totalpol}. In general, any polarization state can be expressed as combination of two Jones vectors $\e_1$, $\e_2$ that represent \emph{orthogonal polarizations}, meaning

\begin{equation}
\label{eq_jones-ortho}
\e^{~}_1 \e^*_2 = 0
\end{equation}

\noindent
where the operator $^*$ denotes complex conjugation.

A (linear) modification of the polarization state of a wave is expressed by a $2\times2$ \emph{Jones matrix} $\J$ that relates input wave $\e$ and output wave $\e'$ like

\begin{equation}
\label{eq_jones-matrix}
\e' = \J \e ~ .
\end{equation}

\noindent
In optics, Jones matrices are commonly employed to characterize polarizing optical elements or trains thereof. Successive modifications $1, 2, ..., n$ of the polarization state can be written in terms of a single Jones matrix. This matrix is given by the product of the matrices corresponding to the individual optical elements like

\begin{equation}
\label{eq_jones-matrix-product}
\J = \J_n\,\J_{n-1}\, ... \,\J_2\,\J_1 ~ .
\end{equation}

\noindent
Simple examples for Jones matrices of polarizing optical elements are

\begin{equation}
\label{eq_jones-matrix-examples}
\J_x = \left[\begin{array}{rr} 1 & 0 \\ 0 & 0 \end{array}\right]; ~~ \J_R = \frac{1}{2} \left[\begin{array}{rr} 1 & i \\ -i & 1 \end{array}\right]
\end{equation}

\noindent
where $\J_x$ describes a linear polarizer with the $x$ axis being the transmission axis, and $\J_R$ corresponds to a right-hand circular polarizer.

\subsubsection{Stokes Parameters \label{sssect_stokes}}

\begin{figure}[t!]
\centering
\includegraphics[trim=0mm 20mm 0mm 5mm,clip,width=82mm]{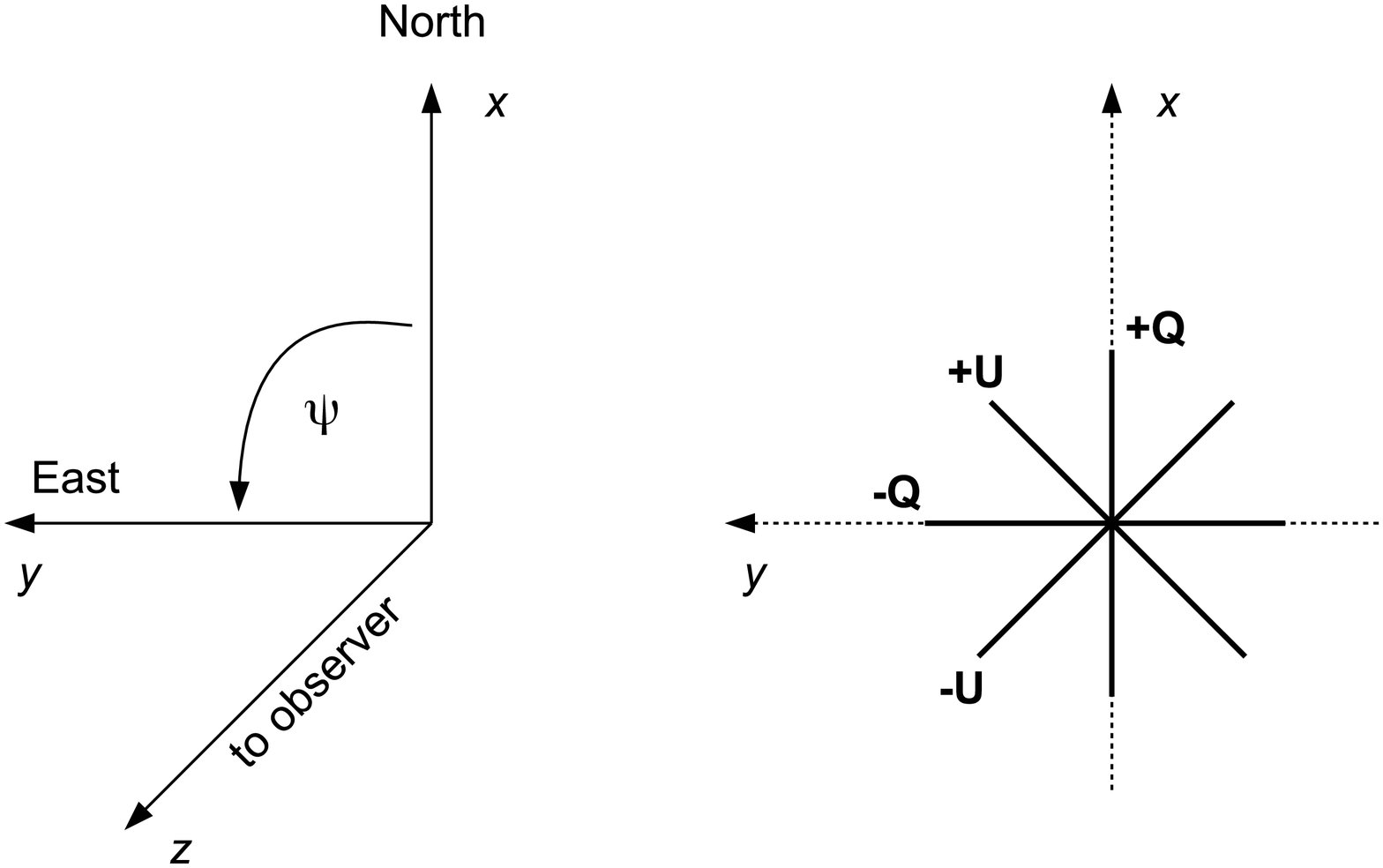}
\caption{Fundamental coordinates and geometries of the Stokes parameters $Q$ and $U$; $\psi$ denotes the parallactic angle. \label{fig_stokes}}
\end{figure}

\noindent
We understand from \S\S\,\ref{sssect_ellpol}--\ref{sssect_circpol} that the polarization state of an electromagnetic wave can be characterized by means of three independent parameters: the amplitudes $E_x(0)$ and $E_y(0)$ and the phase difference $\delta=\phi_2-\phi_1$. In general, astronomical observations deal with light intensities rather than with field amplitudes; accordingly, it is convenient to quantify polarization via characteristic intensities. For reasons that are going to be evident soon, those characteristic intensities\footnote{In the convention we adopt here, intensity $I$ and field amplitude $E$ are related like $I=E^2$, whereas in SI units $I=\varepsilon_0 c E^2$ with vacuum permittivity $\varepsilon_0$. Accordingly, our convention implies a rescaling of electric fields like $E\longrightarrow E'=\sqrt{\varepsilon_0c}\,E$} are the \emph{Stokes parameters}\footnote{Occasionally, the parameters $I$, $Q$, $U$, $V$ are also denoted as $S_0$, $S_1$, $S_2$, $S_3$, respectively.}

\begin{eqnarray}
\label{eq_stokesparams01}
I &=& \al E^2_x \ar + \al E^2_y \ar  \\
Q &=& \al E^2_x \ar - \al E^2_y \ar  \nonumber \\
U &=& 2\,\al E_x E_y \cos\delta \ar  \nonumber \\
V &=& 2\,\al E_x E_y \sin\delta \ar  \nonumber
\end{eqnarray}

\noindent
\citep{stokes1852} where $E_{x,y}\equiv E_{x,y}(t)$ for simplicity, and $\al ... \ar$ denotes the time average of the enclosed parameters taken over times much larger than $2\pi/\omega$ \citep{huard1997,born1999,goldstein2003,thompson2004,wilson2010}. Notably, the parameters $Q$, $U$, and $V$ can take negative values.

By construction, $I$ is the intensity of the wave. The parameter $Q$ quantifies a difference in the intensities in $x$ and $y$, thus providing information on linear polarization. The parameter $U$ quantifies the difference between the two field components diagonal -- at angles of 45$\deg$ and 135$\deg$ counted from the positive $x$ axis -- to the $x$ and $y$ coordinates, thus likewise probing linear polarization. Finally, the parameter $V$ corresponds to the circularly polarized intensity. An illustration of the fundamental geometries is provided in Fig.~\ref{fig_stokes}.

For individual waves -- \emph{microscopic polarization} -- the Stokes parameters are related via

\begin{equation}
\label{eq_stokes-I=Q+U+V}
I^2 = Q^2 + U^2 + V^2 ,
\end{equation}

\noindent
thus reducing the number of free parameters to three -- as expected. As $I$ is a constant, each polarization state of a wave corresponds to a point on a sphere, the \emph{Poincar\'e sphere} \citep{poincare1892}. In case of \emph{macroscopic polarization}, radiation with intensity $I$ is formed from superposition of many elementary emitters; polarization is averaged out at least partially. Accordingly, Eq.~\ref{eq_stokes-I=Q+U+V} breaks down to

\begin{equation}
\label{eq_stokes-Ipol=Q+U+V}
I_P^2 = Q^2 + U^2 + V^2
\end{equation}

\noindent
with $I_P \leq I$ being the polarized intensity; the number of free parameters increases to four, the fourth parameter being $I$. Using the definitions provided by Eqs. \ref{eq_mP}, \ref{eq_mL}, and \ref{eq_mC} as well as the definition of the polarization angle $\chi$, the Stokes parameters relate to the parameters of \emph{linear} polarization like

\begin{eqnarray}
\label{eq_stokes-linpol}
m_L &=& \frac{\sqrt{Q^2 + U^2}}{I} ~~~~~~\, \in [0,1] \\
\chi &=&  \frac{1}{2}\,{\rm atan}_2\left(\frac{U}{Q}\right) ~~~ \in [0,\pi] \nonumber
\end{eqnarray}

\noindent
where atan$_2$ denotes the quadrant-preserving arc tangent; $m_L$ and $\chi$ correspond to the length and the orientation of a vector centered at the origin of a plane spanned by $Q$ and $U$. The degree of \emph{circular} polarization relates to Stokes $V$ like

\begin{equation}
\label{eq_stokes-circpol}
m_C = \frac{V}{I} ~~ \in [-1,1] ~ .
\end{equation}

When using complex exponential notation for the electric field (Eq.~\ref{eq_xyfield}), we find an alternative -- though equivalent -- definition of the Stokes parameters:

\begin{eqnarray}
\label{eq_stokesparams02}
I &=& \al E_x^{~} E_x^* \ar + \al E_y^{~} E_y^* \ar  \\
Q &=& \al E_x^{~} E_x^* \ar - \al E_y^{~} E_y^* \ar  \nonumber \\
U &=& \al E_x^{~} E_y^* \ar + \al E_y^{~} E_x^* \ar  \nonumber \\
V &=& -i \left[ \al E_x^{~} E_y^* \ar - \al E_y^{~} E_x^* \ar \right] ~ . \nonumber
\end{eqnarray}

\noindent
As usual, the operator $^*$ denotes complex conjugation. It is straightforward to see that this definition is equivalent to Eq.~\ref{eq_stokesparams01} (e.g. \citealt{hamaker1996b}).

In astronomy, the Euclidean coordinates we use here are conventionally defined such that the $x$ axis points to the north, the $y$ axis points to the east, and the $z$ axis points toward the observer. Accordingly, the polarization angle $\chi$ is counted from north to east. The orientations of circular polarizations as defined in \S\,\ref{sssect_circpol} -- RHC and LHC -- are preserved.

As I indicated before in \S\,\ref{ssect_photons} and \S\,\ref{sssect_jones}, the polarization state of photons as well as electromagnetic waves can be described using left- and right-hand circular polarization components with amplitudes $E_{L,R}$ and phases $\phi_{L,R}$ as base. Using $\delta'={\phi_R-\phi_L}$, we find (e.g. \citealt{cenacchi2009}) for the Stokes parameters in trigonometric notation

\begin{eqnarray}
\label{eq_stokesparams03}
I &=& \al E^2_R \ar + \al E^2_L  \ar  \\
Q &=& 2\,\al E_R E_L \cos\delta' \ar \nonumber \\
U &=& 2\,\al E_R E_L \sin\delta' \ar  \nonumber \\
V &=& \al E^2_R \ar - \al E^2_L  \ar  \nonumber
\end{eqnarray}

\noindent
and in complex exponential notation

\begin{eqnarray}
\label{eq_stokesparams04}
I &=& \al E_R^{~} E_R^* \ar + \al E_L^{~} E_L^* \ar  \\
Q &=& \al E_R^{~} E_L^* \ar + \al E_L^{~} E_R^* \ar  \nonumber \\
U &=& -i \left[ \al E_R^{~} E_L^* \ar - \al E_L^{~} E_R^* \ar \right] \nonumber \\
V &=& \al E_R^{~} E_R^* \ar - \al E_L^{~} E_L^* \ar ~ . \nonumber
\end{eqnarray}

It is important to note that -- in general -- the treatment of combinations of polarized signals requires the use of the Stokes parameters: only intensities can be added or subtracted in a straightforward manner -- degrees of polarization or polarization angles cannot (cf., e.g., \citealt{heiles2002}).

\subsubsection{M\"uller Formalism \label{sssect_mueller}}

\noindent
The M\"uller formalism \citep{mueller1948,hamaker1996a,huard1997,goldstein2003} extends and combines Jones calculus and Stokes formalism. The Stokes parameters can be expressed as a four-dimensional vector, the \emph{Stokes vector}

\begin{equation}
\label{eq_mueller-stokes}
\SS = \left[\begin{array}{c} I \\ Q \\ U \\ V \end{array}\right] \equiv \T\,\C
\end{equation}

\noindent
with

\begin{equation}
\label{eq_mueller-coherency-transform}
\C = \left[\begin{array}{r} \al E^{~}_xE^{*}_x\ar \\ \al E^{~}_xE^{*}_y\ar \\ \al E^{~}_yE^{*}_x\ar \\ \al E^{~}_yE^{*}_y\ar \end{array}\right] ; ~ \T = \left[\begin{array}{rrrr} 1 & 0 & 0 & 1 \\ 1 & 0 & 0 & -1 \\ 0 & 1 & 1 & 0 \\ 0 & -i & i & 0 \end{array}\right] .
\end{equation}

\noindent
The vector {\C} is commonly referred to as \emph{coherency vector}; evidently, this notation is equivalent to Eq.~\ref{eq_stokesparams02}.

Modifications of the polarization state can be expressed as modifications of the Stokes vector like

\begin{equation}
\label{eq_mueller-stokes-change}
\SS' = \M\,\SS
\end{equation}

\noindent
where {\M} is a $4\times4$ \emph{M\"uller matrix}. Simple examples are

\begin{equation}
\label{eq_mueller-matrix-reflect}
\M_{\rm ref} = \left[\begin{array}{rrrr} 1 & 0 & 0 & 0 \\ 0 & 1 & 0 & 0 \\ 0 & 0 & -1 & 0 \\ 0 & 0 & 0 & -1 \end{array}\right]
\end{equation}

\noindent
and

\begin{equation}
\label{eq_mueller-matrix-rotate}
\M_{\rm rot} = \left[\begin{array}{rrrr} 1 & 0 & 0 & 0 \\ 0 & \cos 2\beta & \sin 2\beta & 0 \\ 0 & -\sin 2\beta & \cos 2\beta & 0 \\ 0 & 0 & 0 & 1 \end{array}\right]
\end{equation}

\noindent
that correspond to a reflection at a mirror and rotation by an angle $\beta$, respectively.

Else than the Jones calculus, the M\"uller formalism can describe unpolarized light as well as depolarization, i.e. a reduction of the degree of (total) polarization $m_P$.

\section{POLARIGENESIS \label{sect_polsources}}  

\noindent
As I discussed briefly in \S\,\ref{sssect_macropol}, the occurrence of a macroscopic polarization of light -- the \emph{polarigenesis} -- is intimately linked to the internal symmetry of the physical system under consideration. Macroscopic polarization requires that the internal structure of the source of light is anisotropic. We can identify a variety of astrophysical sources of polarized light that I discuss in the following.

\subsection{Scattering Polarization \label{ssect_scattering}}

\begin{figure}[t!]
\centering
\includegraphics[width=82mm]{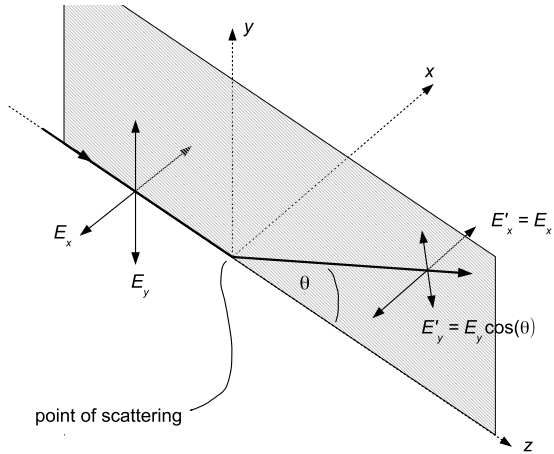}
\caption{The geometry of scattering polarization. An unpolarized light ray with linear polarization components $E_{x,y}$ propagates in positive $z$ direction. The ray is scattered by an angle $\theta$ and continues to propagate in the $yz$ plane (shaded). From the point of view of an observer of the scattered light, the polarization components perpendicular to the direction of propagation are $E'_x=E_x$ and $E'_y=E_y\cos\theta$, respectively: the light is linearly polarized.\label{fig_scatter}}
\end{figure}

\subsubsection{Microscopic Scattering \label{sssect_thomson}}

\noindent
Microscopic scattering processes -- meaning the scattering of a photon at a free electric charge like an electron, an atom, or a molecule -- lead to a characteristic linear polarization of the scattered light. Notably, the following geometry argument holds for a variety of scattering processes like \emph{Thomson scattering}, \emph{Compton scattering}, \emph{Rayleigh scattering}, \emph{fluorescence}, or \emph{Raman scattering} regardless of the different underlying physical mechanisms. From now on, I assume incident unpolarized light propagating in $z$ direction with electric field components $E_{x,y}$ in $x$ and $y$ directions. Upon interaction with a charge, atom, or molecule, the light is scattered and continues to propagate in the $yz$ plane at an angle $\theta$ to the $z$ axis.

From the point of view of an observer of the scattered light, the light intensities in $x$ direction are the same before and after the scattering, i.e. $E'^2_x=E^2_x$, with the prime denoting the scattered light. The $y$ component transforms like $E'^2_y=E^2_y\cos^2\theta$, meaning the intensity is reduced by a factor $\cos^2\theta$. This implies that the scattered light is linearly polarized with a degree of polarization

\begin{equation}
\label{eq_thomson-pol}
m_L = \frac{I'_x - I'_y}{I'_x + I'_y} = \frac{1 - \cos^2\theta}{1 + \cos^2\theta} \in [0,1]
\end{equation}

\noindent
with $I'_{x,y}$ denoting the observed intensities (after the scattering); I provide an illustration in Fig.~\ref{fig_scatter}.

\subsubsection{Scattering by Dust \label{sssect_dust}}

\noindent
Dust grains are an important ingredient of interstellar matter \citep{dyson1997,kwok2007}. Grain sizes are typically on the order of a micrometer, approximately corresponding to the wavelengths of optical to infrared light. A quantitative description of scattering of light by dust is provided by \emph{Mie's theory} that assumes scattering by small spherical particles. From geometrical arguments equivalent to those presented in \S\,\ref{sssect_thomson} one finds that initially unpolarized incident light becomes linearly polarized by dust scattering; the degree of polarization is given by an expression equivalent to Eq.~\ref{eq_thomson-pol} \citep{born1999}.

The derivation of Eq.~\ref{eq_thomson-pol} assumes that the incident light is propagating along a single well-defined direction. In clouds of interstellar matter, this is usually not the case: dust clouds tend to be optically thick, meaning that incident light experiences multiple scattering, absorption, and re-emission events, making the radiation field within the cloud isotropic. In this case, linear polarization can occur if (i) the dust grains have elongated (cylindrical, ellipsoidal) shapes, and (ii) the grains are oriented collectively along a preferred direction by magnetic fields. The absorption of radiation by the dust becomes a function of orientation relative to the magnetic field -- resulting in linear polarization. Empirically, it has been found \citep{serkowski1975,draine2003} that at optical to near-infrared wavelengths the linear polarization is

\begin{equation}
\label{eq_serkowski}
\frac{m_L}{m_L^{\rm max}} \approx \exp\left[ -1.15 \ln^2\left(\frac{\lambda_{\rm max}}{\lambda}\right) \right]
\end{equation}

\noindent
scaled by the degree of linear polarization at a reference wavelength,

\begin{equation}
\label{eq_serkowski-polmax}
m_L^{\rm max} \lesssim 0.03 \, A(\lambda_{\rm max})
\end{equation}

\noindent
where $\lambda$ is the wavelength, $\lambda_{\rm max}\approx550$\,nm, $A$ is the extinction in units of photometric magnitudes, and ln denotes the logarithm to base $e$; this relation is commonly referred to as \emph{Serkowski's law}.

\subsection{Dichroic Media \label{ssect_dichroics}}

\noindent
Dichroism is a further effect of the electric anisotropy of certain materials. Here the attenuation of light due to absorption by the material is anisotropic. Assuming initially unpolarized light, one component of the wave experiences stronger attenuation than the other one; the light becomes polarized.\footnote{The term ``dichroism'' is actually misleading. Historically, the first dichroic crystals studied showed a strong dependence of the effect on the wavelength of the light, leading to rays with different polarization having different colors.} Depending on if the difference in absorption affects the linear or circular wave components, the medium is referred to as \emph{linear dichroic} or \emph{circular dichroic}, respectively. Accordingly, the light becomes either linearly or circularly polarized. The most efficient linear dichroic polarizers are \emph{polaroids}, sheets of organic polymers with long-chain molecules which are aligned by stretching \citep{huard1997,born1999}.

In the presence of a magnetic field, a plasma becomes dichroic with respect to circular polarization. Assuming a propagation of light along the magnetic field lines, the ratio of the absorption coefficients for LHC and RHC polarized light, $\kappa_L$ and $\kappa_R$, respectively, is

\begin{equation}
\label{eq_dichroic-plasma}
\frac{\kappa_L}{\kappa_R} = \left(\frac{\omega + \omega_B}{\omega - \omega_B}\right)^2
\end{equation}

\noindent
where $\omega$ denotes the (angular) frequency of the light and $\omega_B$ denotes the (angular) gyration frequency of charged particles (usually electrons). This relation assumes $\omega_B\ll\omega$; the resulting circular polarization is $m_C=2\omega_B/\omega$ \citep{angel1974}.

\subsection{Optically Active Media \label{ssect_chirality}}

\noindent
Materials composed of helically shaped molecules affect the polarization state of reflected or transmitted light very similar to birefringent and/or dichroic media (\S\S\,\ref{ssect_dichroics}, \ref{sssect_birefringence}). Macroscopic polarization arises if one of the two possible helix orientations is preferred; this is the case in a variety of biological materials. Initially unpolarized light reflected from helically layered surface structures on some insects can reach circular polarizations up to $m_C\approx100$\% \citep{wolstencroft1974}.

\subsection{Synchrotron Radiation \label{ssect_synchrotron}}

\noindent
Synchrotron radiation is arguably the most important type of non-thermal continuum radiation from astronomical sources. It is emitted by electric charges -- usually electrons -- gyrating around magnetic field lines at relativistic velocities. Assuming a magnetic field directed in $z$ direction, the magnetic Lorentz force enforces a circular motion in the $xy$ plane. In addition, the electron will usually have a non-zero velocity in $z$ direction, meaning that the overall trajectory of the electron has a helical shape. From the point of view of an external, not co-moving, observer the radiation is emitted from the electron in forward direction into a narrow cone with half opening angle $\theta\approx1/\gamma$, with $\gamma$ being the relativistic Lorentz factor\footnote{$\,\gamma = \left(1-v^2/c^2\right)^{-1/2}$, with $v$ being the electron speed and $c$ being the speed of light.}\setcounter{footnote}{0} \citep{ginzburg1965,rybicki1979,bradt2008}. Due to geometry, the observer sees the orbit of the electron as an ellipse in projection -- the radiation emitted by a single oscillating charge is thus elliptically polarized.

\begin{figure}[t!]
\centering
\includegraphics[angle=-90,width=82mm]{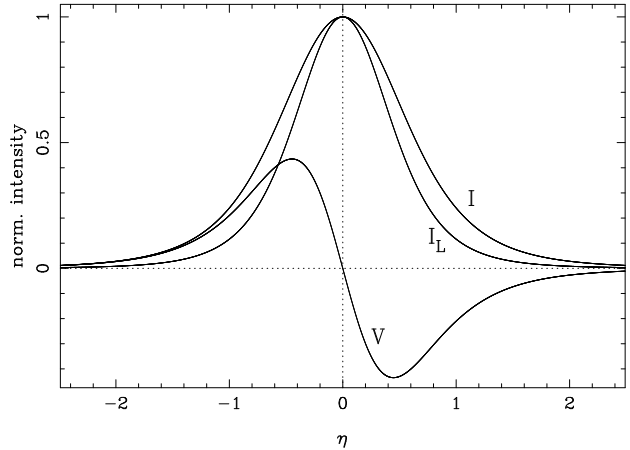}
\caption{Polarization of synchrotron radiation from a collimated electron beam as function of re-scaled viewing angle $\eta=\gamma\sin\phi$. Given here are the Stokes parameters $I$ and $V$ and the linearly polarized flux $I_L=\sqrt{Q^2+U^2}$, normalized to $I(\eta=0)\equiv1$.\label{fig_synchpol}}
\end{figure}

\para{Collimated electron beam.} Macroscopically, the amounts of linear and circular polarization observed by an external observer are functions of $\gamma$ and the viewing angle $\phi$ between the line of sight and the plane of particle motion. For a \emph{collimated} beam of \emph{mono-energetic} electrons -- i.e., all electrons have the same Lorentz factor $\gamma$ -- moving in the $xy$ plane the \emph{circularly} polarized flux (Stokes $V$) is

\begin{equation}
\label{eq_syncpol-V}
V = -\frac{64\eta}{7\pi\sqrt{3}(1+\eta^2)^3}
\end{equation}

\noindent
with $\eta=\gamma\sin\phi$. In this notation, the total intensity (Stokes $I$) of the radiation is given by

\begin{equation}
\label{eq_syncpol-I}
I = \frac{7 + 12\eta^2}{7(1 + \eta^2)^{7/2}}
\end{equation}

\noindent
\citep{michel1991}. The \emph{linearly} polarized flux follows from Eqs.~\ref{eq_syncpol-V} and \ref{eq_syncpol-I} via Eq.~\ref{eq_stokes-I=Q+U+V} in a straightforward manner. Notably, I normalized the expressions for $V$ and $I$ such that $I=1$ at $\phi=0$.

\para{Isotropic electron motion.} In most astrophysical plasmas, the electron velocities are distributed randomly and (more or less) isotropically. This implies that both right-handed and left-handed electron orbits contribute with equal probability, meaning the circular component of the radiation averages out: macroscopically, the synchrotron radiation becomes \emph{linearly} polarized (though not perfectly; see the discussion below).

Taking into account the various projection effects, the degree of linear polarization is given by

\begin{equation}
\label{eq_syncpol-def}
m_L = \frac{I_{\perp}-I_{||}}{I_{\perp}+I_{||}}
\end{equation}

\noindent
with $I_{\perp}$ and $I_{||}$ denoting the intensities perpendicular and parallel to the magnetic field lines as projected onto the plane of the sky. Notably, the direction of polarization is perpendicular to the (projected) direction of the magnetic field. The actual value of $m_L$ depends on the spectrum of the synchrotron radiation which in turn is a function of the distribution of the electron energies. For an ensemble of \emph{mono-energetic} electrons the result is $m_L=75$\% \citep{rybicki1979}. For a \emph{power law distribution} of electron energies, the number of electrons $N$ varies with $\gamma$ like $N\propto\gamma^{-\Gamma}$, with $\Gamma$ being the \emph{energy index}. In this case, the flux density of the synchrotron radiation $S_{\nu}$ varies with frequency $\nu$ like $S_{\nu}\propto\nu^{-\alpha}$, where $\Gamma=2\alpha+1$. The degree of linear polarization is given by

\begin{equation}
\label{eq_syncpol-power}
m_L = \frac{\Gamma+1}{\Gamma+7/3}
\end{equation}

\noindent
\citep{ginzburg1965}. For astrophysically realistic plasmas with $\alpha\approx0...1$, $m_L\approx60...80$\%.

The relation given by Eq.~\ref{eq_syncpol-power} corresponds to a highly idealized situation, assuming optically thin plasmas, isotropic distributions of electrons, perfectly ordered homogeneous magnetic fields, and the absence of substantial perturbations. A major modification occurs for \emph{optically thick} plasmas where each light ray experiences multiple scattering events. In those cases,

\begin{equation}
\label{eq_syncpol-thick}
m_L = \frac{1}{2\Gamma+13/3}
\end{equation}

\noindent
\citep{pachol1970a}; for $\alpha\approx0$, $m_L\approx16$\%. Likewise, any disordering of the magnetic field leads to the polarization signal partially being averaged out, thus reducing $m_L$ well below the idealized theoretical values. In optically thick plasmas, the polarization is oriented \emph{parallel} to the magnetic field component projected on the sky.

Yet another characteristic modulation of polarization is caused by \emph{shocks} propagating through the plasma. The degree of linear polarization in a partially compressed plasma, compared to the case without compression, is reduced by a factor

\begin{equation}
\label{eq_syncpol-compress}
\mu = \frac{\delta}{2-\delta} \in [0,1] ~~ {\rm with} ~~ \delta = (1-k^2)\cos^2\epsilon
\end{equation}

\noindent
where $0\leq k\leq1$ is the factor by which the length of the shocked region is reduced by compression, and $\epsilon$ is the angle between the line of sight and the plane of compression in the frame of reference of the emitter \citep{hughes1985,cawthorne1988}.

The amount of \emph{circular} polarization follows roughly the relation $m_C\sim(\omega/\omega_B)^{-1/2}$ where $\omega$ is the (angular) frequency of the light and $\omega_B$ is the (angular) gyration frequency of the charged particle. As before, this relation assumes a uniform magnetic field and an isotropic distribution of particle velocities. In the extreme case $\omega_B\gtrsim\omega$, the relation changes to $m_C\sim\omega/\omega_B$. In realistic astrophysical plasmas, $m_C\lesssim1$\% \citep{angel1974}.

\subsection{Zeeman Effect \label{ssect_zeeman}}

\noindent
Spectral emission or absorption lines experience modifications if the emitting or absorbing material is permeated by a magnetic field. For atomic or molecular transitions involving the orbital angular momentum only (i.e., no spin--orbit coupling), the quantum mechanical selection rules demand that the change of the magnetic quantum number $m$ has to obey $\Delta m \in [-1,0,1]$ -- meaning there are three different transitions possible. As long as the atom is not exposed to external electric or magnetic fields, the three transitions are energetically degenerate; all three transitions have the same energy and cause spectral lines at the same frequency $\nu_0$. This changes in the presence of an external magnetic field as reported first by \citet{zeeman1897}, hence \emph{Zeeman effect}. As shown by a simple classical analysis \citep{rybicki1979,haken1990}, the magnetic field causes a splitting of the three initially degenerate energy levels. The spectral line splits into a set of three lines located at frequencies $\nu_0$ and $\nu_{\pm}=\nu_0\pm\Delta\nu_{\rm z}$ for $\Delta m=0$ and $\Delta m=\pm1$, respectively. The frequency offset is given by

\begin{equation}
\label{eq_zeeman}
\Delta\nu_{\rm z} = \frac{1}{4\pi}\,\frac{e}{m_{\rm e}}\,B = 14\,{\rm GHz} \times B
\end{equation}

\noindent
where $e$ is the electric charge of the electron, $m_{\rm e}$ is the electron mass, and $B$ is the strength of the magnetic field in units of Tesla. The lines at $\nu_{-}$, $\nu_0$, and $\nu_{+}$ are denoted as $\sigma^{-}$, $\pi$ (``parallel''), and $\sigma^{+}$ components, respectively.

The three spectral lines have distinct polarization properties that depend on the viewing geometry.

\para{Transversal view.} If the line of sight is perpendicular to the magnetic field lines, the observer notes three spectral lines corresponding to the $\pi$ and $\sigma^{\pm}$ components. All lines are \emph{linearly} polarized. The polarization of the $\pi$ component is parallel to the magnetic field (hence the name), the polarizations of the $\sigma^{\pm}$ components are perpendicular to the field lines.

\para{Longitudinal view.} If the line of sight is parallel to the magnetic field lines, only the $\sigma^{\pm}$ components are visible. Both lines are \emph{circularly} polarized, with $\sigma^{+}$ and $\sigma^{-}$ being RHC and LHC, respectively. Notably, the orientation of polarization is defined relative to the direction \emph{of the magnetic field lines}, not the direction of propagation of the light.

Assuming an angle $\theta$ between the magnetic field and the line of sight, the Stokes parameters resulting from the viewing geometry (e.g., \citealt{elitzur2000}) are

\begin{eqnarray}
\label{eq_zeeman-strong-stokes-m0}
I & = & I^0\,\sin^2\theta  \\
Q & = & I^0\,\sin^2\theta  \nonumber \\
U & = & 0  \nonumber \\
V & = & 0 \nonumber
\end{eqnarray}

\noindent
for the $\Delta m=0$ transition and

\begin{eqnarray}
\label{eq_zeeman-strong-stokes-m+-}
I & = & \frac{1}{2}\,I^{\pm}\,(1 + \cos^2\theta)  \\
Q & = & -\frac{1}{2}\,I^{\pm}\,\sin^2\theta \nonumber \\
U & = & 0  \nonumber \\
V & = & \pm I^{\pm} \cos\theta  \nonumber
\end{eqnarray}

\noindent
for the $\Delta m=\pm1$ transitions, respectively. The $I^0$, $I^{\pm}$ denote the maximum intensities of the respective lines, with $I^+(\nu_+) = I^-(\nu_-) = I^0(\nu_0)$.

As yet, I assumed strong Zeeman splitting that leads to distinct spectral lines, meaning intrinsic line widths $\Delta\nu\ll\Delta\nu_{\rm z}$. In astronomy, this is not always the case; the situation $\Delta\nu\gg\Delta\nu_{\rm z}$ is common. In the latter case, the total intensity as function of frequency, $I(\nu)$, exhibits a single line only. The Zeeman components are unveiled by analysis of the Stokes parameters as function of frequency, resulting in

\begin{eqnarray}
\label{eq_zeeman-weak-stokes}
I & = & I^0 + I^+ + I^- = 2\,I^0  \\
Q & = & -\frac{{\rm d}^2I(\nu)}{{\rm d}\nu^2}\,\left(\Delta\nu_{\rm z}\sin\theta\right)^2  \nonumber \\
U & = & 0  \nonumber \\
V & = & \frac{{\rm d}I(\nu)}{{\rm d}\nu}\,\Delta\nu_{\rm z}\cos\theta  \nonumber
\end{eqnarray}

\noindent
Accordingly, $Q(\nu)$ and $V(\nu)$ are the scaled derivatives of $I(\nu)$. Observations especially of $V(\nu)$ are important examples of \emph{spectro-polarimetry}.

As yet, I discussed the \emph{normal} Zeeman effect occurring in atomic transitions without spin--orbit coupling. If spin--orbit coupling has to be taken into account, more complex patterns with multiple line components spaced non-equally occur. Details for this \emph{anomalous} Zeeman effect depend on the total spins $S$, orbital angular momentum quantum numbers $L$, and total angular quantum numbers $J$. The $\pi$ and $\sigma^{\pm}$ components are given by $\Delta M=0$ and $\Delta M=\pm1$ respectively, with $M$ denoting the total magnetic quantum number.

\subsection{Goldreich--Kylafis Effect \label{ssect_goldreich}}

\noindent
We now revisit the Zeeman effect for the case of weak Zeeman splitting ($\Delta\nu\gg\Delta\nu_{\rm z}$), meaning that the three Zeeman line components are not resolved. As indicated by Eq.~\ref{eq_zeeman-weak-stokes}, the linear polarization is rather weak in general (and zero at $\nu=\nu_0$). However, this result is based on the assumption that each transition $\Delta m = -1,0,1$ is equally likely.

In case that the transitions $\Delta m$ occur at \emph{different} rates, increased linear polarization can occur \citep{goldreich1981,goldreich1982}. This is the case if the radiation field within the source is anisotropic: depending on the relative orientation of magnetic field and incident radiation, the amount of anisotropy, and the ratio of collisional and radiative excitation rates, the $\sigma^{\pm}$ and $\pi$ line components are excited with different probabilities. The radiative transition rates ${\mathcal T}$ for the $\sigma^{\pm}$ and $\pi$ components, which are proportional to the line intensities, are given by

\begin{eqnarray}
\label{eq_goldreich}
{\mathcal T}_{\pm} & \propto & \frac{1}{2}\int {\rm d}\Omega\,\left[I_{\perp} + I_{||}\cos^2\alpha\right] \\
{\mathcal T}_{0}   & \propto & \int {\rm d}\Omega\,I_{||}\sin^2\alpha  \nonumber
\end{eqnarray}

\noindent
respectively, where $I_{\perp}$ is the incident light intensity polarized perpendicular to the magnetic field \emph{and} the direction of propagation, $I_{||}$ is the intensity of the incident light polarized perpendicular to $I_{\perp}$ \emph{and} the direction of propagation, $\Omega$ is the solid angle, and $\alpha$ is the angle between the travel path of the incident light and the magnetic field.

Depending on which transitions are excited preferentially, the resulting linear polarization is oriented either parallel or perpendicular to the (sky-projected) magnetic field lines. Under realistic conditions, one may expect degrees of linear polarization up to $\approx$10\% in molecular lines from cool (temperatures $\lesssim$100\,K) interstellar matter at millimeter-radio wavelengths.

\subsection{Hanle Effect \label{ssect_hanle}}

\noindent
The Hanle effect \citep{hanle1924} is a phenomenon that appears in fluorescent light. In the following, I assume a fluorescent gas permeated by a weak magnetic field $B$. The primary source of radiation is located in $x$ direction from the fluorescent gas, the line of sight as well as the magnetic field are directed along the $z$ axis. The primary light source emits radiation in $x$ direction with electric field components $E_{y,z}$. By geometry, the linear polarization component $E_y$ causes fluorescence observable in $z$ direction.

When regarding normal Zeeman splitting (\S\,\ref{ssect_zeeman}) in the limit of a vanishing magnetic field $B\rightarrow0$, one reaches the regime of \emph{coherent} resonances. As long as the Zeeman splitting is on the order of the natural line width, i.e., $\Delta\nu_z\gtrsim\Delta\nu$, the $\sigma^{\pm}$ transitions are excited independently. The observer notes two circularly polarized waves with amplitudes $E_+$ and $E_-$ and a combined intensity $I=E_+^2+E_-^2$. For vanishing magnetic fields, $\Delta\nu_z\ll\Delta\nu$ and both transitions $\sigma^{\pm}$ can be excited by the \emph{same} photon (which is linearly polarized and thus can be decomposed in one RHC and one LHC component for exciting $\sigma^+$ and $\sigma^-$, respectively). The observer notes fluorescent light with intensity $I=(E_++E_-)^2$ and linear polarization ($m_L=100$\% in ideal situations) directed along $y$.\footnote{Similar situations can also occur at non-zero magnetic fields where magnetic term levels belonging to \emph{different} angular momentum quantum numbers can cross. This is the base of \emph{level-crossing spectroscopy}.}

Ideal coherence with maximum light intensity $I$ and maximum linear polarization occurs at $B=0$. For an increasing field strength ($|B|>0$), the degree of coherence decreases, causing several effects: (i) the intensity of the fluorescent light decreases; (ii) the ``de-phasing'' of $\sigma^+$ and $\sigma^-$ causes (a) a depolarization, and (b) a turning of the plane of linear polarization by an angle $\beta$ such that $\tan\beta\approx\omega_L/\Gamma_{\rm line}$, with $\omega_L$ being the \emph{Lamor frequency} and $\Gamma_{\rm line}$ being the effective (i.e., natural plus collisional) line width (e.g., \citealt{stenflo1982}).

\subsection{Interactions of Polarization and Matter \label{ssect_matter}}

\subsubsection{Partial Reflection \label{sssect_reflection}}

\noindent
A plane wave that falls onto the boundary between two homogeneous media ``1'' and ``2'' with indices of refraction $n_{1,2}$ -- like, e.g., the boundary between two plane-parallel plates made of different types of glass -- is split into a reflected and a transmitted wave according to the law of reflection and Snell's law of diffraction, respectively. Except of the special case of normal incidence, reflectivity $R$ and transmissivity $T$ -- the fractions of intensity being reflected and transmitted -- are functions of linear polarization \citep{born1999}. Both parameters have to be re-written like

\begin{eqnarray}
\label{eq_reflection-components}
R &=& R_{||}\cos^2\alpha + R_{\perp}\sin^2\alpha  \\
T &=& T_{||}\cos^2\alpha + T_{\perp}\sin^2\alpha \nonumber
\end{eqnarray}

\noindent
with $||$ and $\perp$ denoting linear polarizations parallel and perpendicular to the plane of incidence, respectively, and $\alpha$ being the angle between the electric field vector and the plane of incidence. Conservation of energy demands

\begin{equation}
\label{eq_reflection-norm}
R + T = R_{||} + T_{||} = R_{\perp} + T_{\perp} = 1 ~ .
\end{equation}

All reflectivities and transmittivities depend on the ratio $n = n_2/n_1$ and the angle between the normal of the boundary and the incident light ray, $\theta$, via \emph{Fresnel's formulae}. In general, $R_{||} \neq R_{\perp}$ and $T_{||} \neq T_{\perp}$; both, reflected and transmitted light become linearly polarized even if the incident light is unpolarized. In the specific case

\begin{equation}
\label{eq_reflection-brewster}
\tan\theta_B = n
\end{equation}

\noindent
the component $R_{||}$ vanishes; $\theta_B$ is the \emph{Brewster angle}. For the case of a transition from air to glass, $n\approx1.5$ and $\theta_B\approx57\deg$. If the incident light is unpolarized, the \emph{reflected} light is completely polarized; the \emph{transmitted} light is polarized with a degree of linear polarization $m_L\approx8$\%.

\subsubsection{Birefringence \label{sssect_birefringence}}

\noindent
Birefringence is a consequence of the \emph{electric anisotropy} of crystals, meaning that the response of the medium to incident radiation (usually) depends on the direction of the electric field. This anisotropy is described by a symmetric \emph{dielectric tensor} {$\bf\varepsilon$} that defines a system of \emph{principal dielectric axes} with permittivities $\{\varepsilon_x,\varepsilon_y,\varepsilon_z\}$ \citep{fowles1975,huard1997,born1999}. In case of \emph{isotropic} crystals, $\varepsilon_x=\varepsilon_y=\varepsilon_z$; this is the case for cubic crystals. In \emph{uniaxial} crystals, $\varepsilon_x=\varepsilon_y$ while $\varepsilon_y\neq\varepsilon_z$; in \emph{biaxial} crystals, $\varepsilon_x\neq\varepsilon_y\neq\varepsilon_z$. The two linear polarization components of a light ray -- which are perpendicular to the direction of propagation -- passing through an anisotropic crystal experience different permittivities and thus different indices of refraction $n\propto\sqrt{\varepsilon}$. The \emph{optic axes} of a crystal correspond to light travel paths for which the two linear polarizations experience equal refraction. For the case of uniaxial crystals, the optic axis is the $z$ axis.

Birefringence is employed in various optical elements that modify the polarization state of light.

\para{Polarizers.} For a uniaxial crystal, the dielectric tensor is an ellipsoid. The index of refraction describes a circle in the $xy$ plane and ellipses in the $xz$ and $yz$ planes,with the optic axis corresponding to either the major or the minor axis of the ellipse depending on the material. For a wave polarized perpendicular to the optic axis, $E_1$, the propagation in the crystal is isotropic, the surfaces of equal phase in the plane of incidence are circular (regardless of the relative orientation of optic axis and plane of incidence). $E_1$ propagates through the crystal according to Snell's law; therefore it is referred to as \emph{ordinary wave}. For a wave $E_2$ polarized perpendicular to $E_1$ (i.e., $E_1\perp E_2$), the components parallel and perpendicular to the optic axis experience different refractive indices and thus different phase velocities. The surfaces of equal phase in the plane of incidence are elliptical, rendering Snell's law invalid; accordingly, $E_2$ is referred to as \emph{extraordinary wave}. Anisotropic crystals can be used to separate the ordinary and extraordinary waves, and thus the two linear polarizations, of unpolarized light. Various crystal geometries and combinations of crystals with different optic axis orientations -- adding internal reflections -- are used. The most common types of polarizers are \emph{polarizing beam splitters}, \emph{Nicol prisms}, and \emph{Glan-Thompson prisms}. 

\para{Linear-to-circular converters.} Let us consider a thin, plane-parallel plate cut from a uniaxial crystal located in the $xy$ plane, with the optic axis being the $y$ axis. In this case, the linear polarization components of a light wave propagating in $z$ direction experience two different refractive indices $n_{x,y}$. The corresponding phase velocities -- the speeds of light within the crystal -- are $c_x=c/n_x$ and $c_y=c/n_y$, respectively, with $c$ denoting the speed of light in vacuum. For light with wavelength $\lambda$ in vacuum, traveling a distance $d$ within the crystal, the two polarizations experience a phase shift

\begin{equation}
\label{eq_quarterwave-phaseshift}
\delta = \frac{2\pi}{\lambda}\,d\,(n_y - n_x) ~ .
\end{equation}

\noindent
When choosing $d$ such that $d(n_y-n_x)=\lambda/4$, we find $\delta=\pi/2$. Light with $\al E_x^2\ar = \al E_y^2\ar$, i.e. Stokes parameter $Q=0$ (cf. Eq.~\ref{eq_stokesparams01}), becomes circularly polarized (cf. Eq.~\ref{eq_circpolwaves}); for $Q\neq0$, it becomes elliptically polarized. Optical elements with this property are referred to as \emph{quarter wave plates}. For realistic uniaxial crystals with $|n_y-n_x|\approx0.01...0.1$ \citep{fowles1975}, $d\approx(3...25)\lambda$, meaning that quarter wave plates are fragile devices.

\para{Polarization plane turners.} The relation provided by Eq.~\ref{eq_quarterwave-phaseshift} has an additional consequence for linearly polarized light with components $E_{x,y}$. When choosing $d$ such that $d(n_y-n_x)=\lambda/2$, the phase shift becomes $\delta=\pi$; the plate is a \emph{half wave plate}. The component perpendicular to the optic axis, here $E_x$, is mirrored at the optic axis, i.e. $E_x\longrightarrow-E_x$ (cf. Eq.~\ref{eq_linpolwaves}). If plane of polarization and optic axis are tilted by an angle $\beta$, the crystal turns the polarization plane of the light by an angle of $2\beta$. 

In a variety of materials, birefringence can be induced by external electric -- \emph{Kerr effect} -- or magnetic -- \emph{Cotton-Mouton effect} -- fields \citep{fowles1975}. This is employed in light modulators that need to be switched between different states at high speed. In case of the Kerr effect, the difference between the indices of refraction parallel -- $n_{||}$ -- and perpendicular -- $n_{\perp}$ -- to the orientation of an external electric field with amplitude $E$ is

\begin{equation}
\label{eq_kerr}
n_{||} - n_{\perp} = K\,E^2\,\lambda
\end{equation}

\noindent
where $\lambda$ is the wavelength of the light in vacuum and $K$ is \emph{Kerr's constant} which is a function of the material. The Cotton-Mouton effect is the magnetic analogue of the (electric) Kerr effect; here the difference between the two indices of refraction is proportional to the squared strength of the external \emph{magnetic} field.

A further variety is introduced by the \emph{Pockels effect} observed in certain kinds of birefringent crystals upon application of external electric fields. Here the difference between the two indices of refraction is proportional to the electric field strength. This effect is used in \emph{Pockels cells} that permit a rapid modulation of light. A common setup comprises a Pockels cell located between two static linear polarizers with perpendicular transmission axes. Via appropriate switching of the Pockels cell, it turns the plane of polarization of the infalling linearly polarized light, making the setup act as a very fast shutter \citep{fowles1975}.

\subsubsection{Faraday Rotation \label{sssect_faraday-rot}}

\noindent
An external, static magnetic field $\B$ permeating a medium introduces an electric anisotropy. The impact on a light wave propagating through the medium is found by solving the equation of motion for an electron influenced by $\B$ and the oscillating electric field of the light wave $\E(t)$ \citep{fowles1975,rybicki1979}. From this, one finds a circular electric anisotropy with permittivities $\varepsilon_R\neq\varepsilon_L$, with $R$ and $L$ denoting right-hand and left-hand circular polarization, respectively. As any linearly polarized wave can be expressed as a superposition of one left-hand and one right-hand circularly polarized wave (Eq.~\ref{eq_circ-lin}), the electric anisotropy implies a characteristic rotation of the plane of polarization of linearly polarized light. The change of polarization angle can be expressed like

\begin{equation}
\label{eq_faraday-solid}
\Delta\chi = {\mathcal{V}}\,B_{||}\,l
\end{equation}

\noindent
with $l$ denoting the length of the light travel path within the medium, $B_{||}$ being the magnetic field strength parallel to the light travel path, and $\mathcal{V}$ denoting \emph{Verdet's constant} which is a function of wavelength and material \citep{fowles1975}.

In astrophysical situations, Faraday rotation occurs when light passes through magnetized interstellar plasma. This effect is quantified like

\begin{equation}
\label{eq_faraday-plasma}
\Delta\chi = {\rm RM}\times\lambda^2
\end{equation}

\noindent
where $\lambda$ is the wavelength of the radiation (in the rest-frame of the medium) and RM is the \emph{rotation measure} (in units of $\rm rad\,m^{-2}$)

\begin{equation}
\label{eq_rotationmeasure}
{\rm RM} = 8.1\times10^5 \int_{0}^{l} \, B_{||} \, n_e \, {\rm d}z
\end{equation}

\noindent
with $B_{||}$ being the strength of the magnetic field (in units of Gauss) parallel to the line of sight (l.o.s.), $n_e$ being the electron number density (in cm$^{-3}$), and $z$ being the coordinate (in parsec) directed along the l.o.s. \citep{rybicki1979,wilson2010}.

\subsubsection{Faraday Depolarization \label{sssect_faraday-depol}}

\noindent
In case of spatially inhomogeneous media, especially astrophysical plasmas, the Faraday effect can lead to a loss of linearly polarized intensity. If the rotation measure RM shows modulations with amplitudes $\Delta$RM on spatial scales smaller than the source, the source radiation experiences different Faraday rotation depending on the position. Observations that do not resolve the RM structure spatially superimpose waves with different orientations of their planes of linear polarization. This partially averages out the polarization signal, reducing the degree of linear polarization observed. A complete depolarization occurs when the medium is ``Faraday thick''; from Eq.~\ref{eq_faraday-plasma} one can estimate that this is the case if

\begin{equation}
\Delta{\rm RM}\times\lambda^2 \gg 1 ~~ .
\label{eq_faraday-depth}
\end{equation}

\noindent
A more sophisticated calculation is possible when assuming that the RM fluctuations follow a Gaussian distribution with dispersion $\zeta\approx\Delta$RM. For a source that is not resolved spatially by observations, one finds a depolarization law

\begin{equation}
\xi = \exp\left(-2\zeta^2\lambda^4\right)
\label{eq_faraday-depol}
\end{equation}

\noindent
\citep{burn1966,tribble1991}. The parameter $\xi\in[0,1]$ is the ratio of observed and intrinsic degree of linear polarization.

\subsubsection{Polarization Conversion \label{sssect_polconversion}}

\noindent
Under certain conditions, effects corresponding to those of birefringence in crystals can be observed also in astrophysical plasmas. In the following, I assume an electromagnetic wave with components $E_{x,y}$ propagating through a plasma in $z$ direction. The plasma is permeated by an ordered, static magnetic field directed along the $x$ axis. Plasma electrons accelerated by $E_x$ can move freely, whereas those accelerated by $E_y$ experience an additional magnetic Lorentz force -- the response of the plasma to light becomes anisotropic, the plasma effectively becomes birefringent. In analogy to the relation given by Eq.~\ref{eq_quarterwave-phaseshift}, this effective birefringence introduces a phase shift between $E_x$ and $E_y$ that converts linear into circular polarization and vice versa; this effect is also referred to as \emph{Faraday conversion} or \emph{Faraday pulsation} \citep{pachol1970}. 

In relativistic astrophysical plasmas and at radio frequencies, one may expect to observe a certain level of circularly polarized light generated from initially linearly polarized radiation. Details depend strongly on the physical conditions within the plasma. \citet{pachol1973} provides an estimate for the relation between the degrees of linear ($m_L$) and circular ($m_C$) polarization,

\begin{equation}
\label{eq_polconversion}
\frac{m_C}{m_L} \propto n_e \, B_{\perp}^2 \, \nu^{-3}
\end{equation}

\noindent
where $n_e$ denotes the electron density, $B_{\perp}$ is the strength of the magnetic field component perpendicular to the line of sight, and $\nu$ is the observing frequency. In general, one may expect $m_C\lesssim1$\%.

\subsubsection{Chandrasekhar-Fermi Effect \label{sssect_chandra}}

\noindent
Linear polarization generated within a magnetized turbulent plasma -- via, e.g., dust scattering (\S\,\ref{sssect_dust}) or synchrotron radiation (\S\,\ref{ssect_synchrotron}) -- is sensitive to the strengths of turbulence and magnetic field. The magnetic field is assumed to be ``frozen'' in the plasma. In case of weak fields, the field lines are dragged around by the turbulence, leading to a large r.m.s. dispersion in polarization angles. In case of strong fields, the field lines remain rather unimpressed by the turbulence, the dispersion in polarization angles is small. Magnetic field, turbulence, and polarization angle are related \citep{chandra1953} like

\begin{equation}
\label{eq_chandra}
B_{\perp} = \left(\frac{4}{3}\pi\rho\right)^{1/2} \frac{\sigma_v}{\sigma_{\chi}}
\end{equation}

\noindent
where $B_{\perp}$ is the strength of the magnetic field perpendicular to the line of sight (in Gauss), $\rho$ is the mass density of the gas (in g\,cm$^{-3}$), $\sigma_v$ is the r.m.s. velocity dispersion of the gas (in cm\,s$^{-1}$), and $\sigma_{\chi}$ is the dispersion of polarization angles (in radians).

\section{OBSERVATIONS \label{sect_observe}}  

\noindent
Similar to the cases of photometry and spectroscopy, the techniques used for polarimetry of radiation from astronomical sources depend strongly on the energy of the light. In general, we can distinguish three different wavelength regimes. At \emph{radio} wavelengths, we are able to record electromagnetic waves with their amplitudes and phases. At \emph{optical} wavelengths, we are usually dealing with light intensity information. At \emph{X/$\gamma$-ray} energies, a combination of high frequency and low flux usually implies that we observe and analyze individual photons. This being said, I note that the distinction can be blurred depending on the physical situation, and various techniques find application over a wide range of radiation energies.

\subsection{Sky Projection \label{ssect_sky}}

\noindent
Our usual use of polarization parameters, especially of the Stokes parameters (\S\,\ref{sssect_stokes}), implicitly assumes that emitter and receiver of radiation are placed in a common, stationary system of coordinates. In astronomical observations, this is usually not the case: Earth rotation leads to a rotation of the field of view with respect to the observer. Assuming electric fields $E_{\rm V,H}$ measured vertical and horizontal, respectively, with respect to the telescope, these are related to the Stokes parameters in the frame of reference of the source on sky like

\begin{eqnarray}
\label{eq_projection}
2\,\al E_{\rm V}E_{\rm V}^*\ar &=& I + Q\cos2\psi + U\sin2\psi    \\
2\,\al E_{\rm H}E_{\rm H}^*\ar &=& I - Q\cos2\psi - U\sin2\psi    \nonumber \\
2\,\al E_{\rm V}E_{\rm H}^*\ar &=& -Q\sin2\psi + U\cos2\psi + iV  \nonumber \\
2\,\al E_{\rm H}E_{\rm V}^*\ar &=& -Q\sin2\psi + U\cos2\psi - iV  \nonumber
\end{eqnarray}

\noindent
where $\psi$ denotes the \emph{parallactic angle} counted from north to east and $i$ is the imaginary unit. Comparison to Eq.~\ref{eq_stokesparams02} shows that Earth rotation leads to a conversion from $Q$ to $U$ and vice versa from the point of view of the observer; $V$ remains unaffected \citep{thompson2004}. The same result follows from Eq.~\ref{eq_mueller-matrix-rotate} in a straightforward manner.

\subsection{Terrestrial Atmosphere \label{ssect_atmosphere}}

\noindent
In most situations, the influence of Earth's atmosphere on polarization can be neglected; the atmosphere is neither birefringent nor dichroic. An important exception occurs at radio frequencies where the interaction of ionosphere and terrestrial magnetic field causes substantial Faraday rotation (\S\,\ref{sssect_faraday-rot}). At an observing frequency $\nu=100$\,MHz, the angle of polarization is rotated by $\Delta\chi\approx300\deg$ at night to $\Delta\chi\approx3\,000\deg$ at daytime under typical atmospheric conditions; a reliable derivation of the true polarization angle is very difficult. As $\Delta\chi\propto\nu^{-2}$, this effect can be circumvented by selecting a sufficiently high observing frequency \citep{thompson2004,clarke2010}.

An additional effect relevant mostly at optical wavelengths is the polarization of scattered sun or moon light. From the geometry argument presented in \S\,\ref{sssect_thomson} it is straightforward to see that scattered light is linearly polarized. The degree of polarization reaches its maximum at an angular distance of $90\deg$ from the light source, the polarization is oriented perpendicular to the line on sky connecting the source and the point observed. This behavior can be exploited for the calibration of polarimetric observations via dedicated observations of scattered light.

\subsection{Instrumental Polarization \label{ssect_instrument}}

\noindent
The design and geometry of a telescope inevitably influence the polarization of the collected light. Except of highly symmetric situations -- most notably in Cassegrain focus telescopes -- the (usually) multiple reflections within an optical system alter the polarization state of the light. The resulting \emph{instrumental polarization} is given by the product of the M\"uller matrices of the individual telescope components as described in \S\,\ref{sssect_mueller}. These geometric effects need to be corrected in the course of data analysis and/or by dedicated corrective optics in the telescope (see, e.g., \citealt{thum2008} for a discussion of Nasmyth optics).

Even though one may correct for the influence of the telescope geometry, realistic instruments are not perfect. In the most general case, the observed Stokes parameter values deviate from the actual ones by

\begin{equation}
\label{eq_errormatrix}
\small
\Delta\SS = -\frac{1}{2}\left[\begin{array}{rrrr} \gamma_{++} & \gamma_{+-} & \delta_{+-} & -i\delta_{-+} \\ \gamma_{+-} & \gamma_{++} & \delta_{++} & -i\delta_{--} \\ \delta_{+-} & -\delta_{++} & \gamma_{++} & i\gamma_{--} \\ -i\delta_{-+} & i\delta_{--} & -i\gamma_{--} & \gamma_{++} \end{array}\right]\SS
\end{equation}

\noindent
where $\SS$ is the Stokes vector of the infalling light, the $\gamma_{\rm xx}\ll1$ are error terms related to the \emph{gains}, or efficiencies, of the optical paths for the $E_{x,y}$ components, and the $\delta_{\rm xx}\ll1$ are error terms related to the \emph{leakage}, also known as \emph{cross-talk}, meaning the mutual influence of the optical paths for separate polarizations;\footnote{Even though the $\delta$ and $\gamma$ terms are small, they can easily be of the order of few per cent -- i.e. the same order as the actual polarization signal in many cases.} the (maximum) number of error terms is seven \citep{sault1996}. The expression given by Eq.~\ref{eq_errormatrix} assumes that Stokes parameters are derived from linear polarization components via Eq.~\ref{eq_stokesparams02}. If Stokes parameters are derived from circular polarization components (Eq.~\ref{eq_stokesparams04}), Eq.~\ref{eq_errormatrix} can be applied to a Stokes vector with $Q$, $U$, and $V$ being interchanged with $V$, $Q$, and $U$, respectively (cf. Eq.~\ref{eq_stokesparams02} vs. Eq.~\ref{eq_stokesparams04}).

The actual calibration procedure depends strongly on the telescope(s) used. In general, calibration involves observations of one or more \emph{unpolarized} astronomical reference sources -- probing interactions between $I$ on the one hand and $Q$, $U$, and $V$ on the other hand -- and observations of one or more \emph{polarized} calibration sources, possibly several times at different parallactic angles -- probing interactions between $Q$, $U$, and $V$ by comparison of observed and expected values (e.g., \citealt{clarke2010}).

\subsection{Polarization Statistics \label{ssect_statistics}}

\noindent
Whereas the effects discussed in \S\S\,\ref{ssect_sky}--\ref{ssect_instrument} introduce \emph{systematic} errors into polarization data, we now discuss the \emph{statistical} uncertainties and limits to be taken into account. First of all, it is important to note that, in general, polarimetric observations require much better signal-to-noise ratios ($S/N$) than photometric ones. As the degrees of linear or circular polarization $m_{L,C}\leq1$ are usually much smaller than unity, the signal-to-noise ratio $(S/N)_I$ of the total intensity (Stokes $I$) signal can be related to the $S/N$ of the polarized intensity $(S/N)_P$ like $(S/N)_P \approx m_{L,C}(S/N)_I$.\footnote{This relation is strictly valid only when the polarization is derived from sums or differences of intensities, especially in optical polarimetry. In cases where the polarization is derived from multiplications of fields or from correlations, the process of multiplication leads to non-Gaussian error distributions, modifying the noise estimates by factors of several.} This implies that a detection of a weak polarization signal may require very high $(S/N)_I$.

In case of linear polarization, statistical measurement uncertainties lead to a \emph{bias} in the measured values for $m_L$. This is due to $m_L$ being positive definite by construction (Eq.~\ref{eq_stokes-linpol}): even if $Q$ and $U$ are symmetric random variables centered at zero, the sum of their squares is not; the values of $m_L$ follow a \emph{Rice distribution}. A de-biasing can be attempted by subtracting from each of $Q$ and $U$ the corresponding statistical uncertainty in squares before the calculation of $m_L$. For high $(S/N)_P$, the statistical errors of $m_L$, $\sigma_m$, and polarization angle $\chi$, $\sigma_{\chi}$, are related like $\sigma_{\chi}=\sigma_m/(2m_L)$ (in units of radians); the values of $\chi$ follow a normal distribution. For low $(S/N)_P$, the values for $Q$ and $U$ scatter around the origin of the $QU$ plane, the distribution of the $\chi$ values becomes more and more platykurtic for lower and lower $(S/N)_P$ \citep{clarke2010}.

\subsection{Radio Observations \label{ssect_radio}}

\noindent
At radio wavelengths, the infalling radiation can be recorded and analyzed as waves with full amplitude and phase information; due to fundamental quantum limits, this is possible at frequencies up to about one THz \citep{thompson2004,wilson2010}. Regardless of the actual design details, a radio telescope can be modeled as a cross of two dipoles aligned along the $x$ and $y$ axes, respectively. We may assume, as usual, light propagating along the $z$ direction with linear polarization components $E_{x,y}$. Each of the two dipoles receives the corresponding polarization component and converts it into an electric voltage that can be recorded and processed electronically -- radio receivers are polarimeters by construction\footnote{This excludes radio techniques sensitive to total intensities only, notably bolometers. These have to be treated like optical telescopes (\S\,\ref{ssect_optical}).} (see also \citealt{hamaker1996a,hamaker1996b,sault1996,hamaker2000,hamaker2006} for an exhaustive discussion). The signals received by the dipoles can be \emph{auto}correlated -- resulting in the time-averaged products $\al E_xE_x^*\ar$ and $\al E_yE_y^*\ar$  -- as well as \emph{cross}correlated -- resulting in $\al E_xE_y^*\ar$ and $\al E_yE_x^*\ar$ (using complex exponential notation). The Stokes parameters $I, Q, U, V$ are derived from these products via Eq.~\ref{eq_stokesparams02} in a straightforward manner. A receiving system sensitive to both polarizations $E_{x,y}$ (or $E_{R,L}$  for circular polarization) is referred to as a \emph{dual-polarization receiver}. 

The cross of dipoles also serves as a model for radio receivers sensitive to \emph{circular} polarization. For this we assume that (i) the signals from the two dipoles are sent to a common electronic processor and summed up coherently and (ii) a phase shift of $\pm\pi/2$ is applied to the signal from the $y$ dipole. Accordingly, the two voltages will be in phase and trigger a signal if the infalling light is either RHC or LHC polarized, depending on the sign of the phase shift. The receiver is a \emph{single-polarization receiver} sensitive to either RHC or LHC; it can be extended to a dual-polarization receiver by adding a second cross of dipoles with opposite phase shift. By symmetry, these arguments hold also for the case of \emph{sending} waves for \emph{radar astronomy}; here, usually circularly polarized radio light is used \citep{ostro1993}. Due to the technical simplicity of radio polarimetry, recent efforts have been directed toward simultaneous multi-wavelength polarimetry (cf., e.g., \citealt{kim2011,sslee2011}) aimed at the measurement of differential parameters like dispersion measures.

The choice of polarization is important when combining the signals from two antennas located at large distance, e.g., in Very Long Baseline Interferometry (VLBI). The discussion provided in \S\,\ref{ssect_sky} also implies that the \emph{raw observed} values for $Q$ and $U$ are functions of the geographic positions when using linear polarization receivers. This problem is circumvented by using circular polarization receivers which are insensitive to Earth rotation.

\subsection{Optical Observations \label{ssect_optical}}

\begin{figure}[t!]
\centering
\includegraphics[trim=0mm 25mm 0mm 10mm,clip,width=82mm]{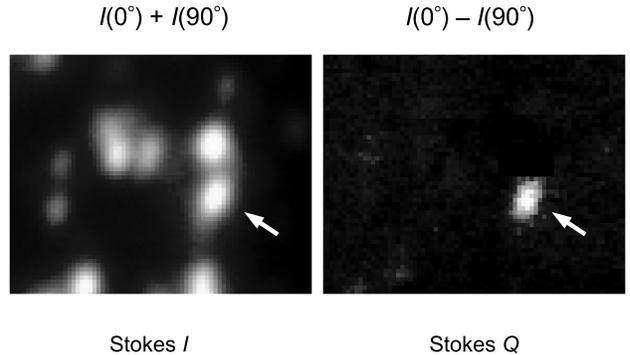}
\caption{Polarimetric imaging of Sagittarius~A* (Sgr~A*), the supermassive black hole at the center of the Milky Way, at 2.2\,$\mu$m. Sgr~A* is indicated by an arrow, the surrounding sources are stars; the angular resolution is $\approx$50\,mas. The difference between the synchrotron source Sgr~A* and the stars emitting thermal radiation becomes evident in the Stokes $Q$ image; at the time of observation, $m_L\approx20$\% \citep{trippe2007}.\label{fig_optpol}}
\end{figure}

\noindent
At optical wavelengths, polarimetry is limited to light intensities rather than electric waves. Polarimetric measurements require the use of polarizers plus auxiliary optical elements placed in the optical path before the detector (usually a CCD array; e.g., \citealt{tinbergen1996}).

\emph{Linear} polarization can be probed by measuring the intensity of the received light, $I(\psi)$, polarized at a parallactic angle $\psi$ to the $x$ (north-south) axis of the usual $xy$ coordinate system of the Stokes parameters (\S\,\ref{sssect_stokes}). The Stokes parameters $Q$ and $U$ are related to these intensities like

\begin{eqnarray}
\label{eq_optlinpol}
\frac{Q}{I} &=& \frac{I(0\deg) - I(90\deg)}{I(0\deg) + I(90\deg)}  \\
\frac{U}{I} &=& \frac{I(45\deg) - I(135\deg)}{I(45\deg) + I(135\deg)}  \nonumber
\end{eqnarray}

\noindent
with $I\equiv I(0\deg) + I(90\deg)\equiv I(45\deg) + I(135\deg)$ being Stokes $I$ as usual \citep{kitchin2009,witzel2011}; see Fig.~\ref{fig_optpol} for an example. Alternatively, one may measure $I(\psi)$ at multiple -- at least four -- values of $\psi$ and model the measurement values with the function

\begin{equation}
\label{eq_rotpol}
q(\psi) = \frac{I(\psi)-I(\psi+90\deg)}{I(\psi)+I(\psi+90\deg)} = m_L\cos\left[2(\psi-\chi)\right]
\end{equation}

\noindent
with $m_L$ denoting the degree of linear polarization and $\chi$ denoting the polarization angle as defined in \S\,\ref{sssect_macropol} (e.g., \citealt{ott1999,trippe2010}). Using Eq.~\ref{eq_rotpol} with a sufficiently large number of measurement values ($\geq$8) with a good sampling of $\psi$ values helps to recognize instrumental polarization effects in the data.\footnote{This is straightforward to see in the special case of Cassegrain focus observations. In this case, the target polarization is fixed with respect to the sky whereas the instrumental polarization is fixed with respect to the telescope. We may obtain observations at two (or more) different hour angles and model each data set as a superposition of two cosine profiles as defined in Eq.~\ref{eq_rotpol}: one corresponding to the target polarization and one corresponding to the instrumental polarization. A polarization signal which remains unchanged -- in sky coordinates -- at different hour angles is intrinsic to the target.}

In order to filter $I(\psi)$ out of the infalling radiation, various types of polarizers can be used. One possibility are \emph{wire-grid polarizers} \citep{huard1997} that pass light polarized perpendicular to the grid and reflect the other component. For each value of $\psi$, the grid is rotated into the required position and an image of the target is taken (e.g., \citealt{ott1999}). A more efficient approach is provided by using a combination of (i) a \emph{Wollaston prism} that splits the infalling light into ordinary and extraordinary linearly polarized rays, and (ii) a \emph{half wave plate} (HWP) that permits turning the plane of linear polarization (cf. \S\,\ref{sssect_birefringence}). A complete measurement cycle involves taking two images, each showing the ordinary and extraordinary ray images of the target: one with the HWP turned to a position corresponding to $\psi=0\deg/90\deg$, one with the HWP turned such that $\psi=45\deg/135\deg$ is observed. The linear polarization of the target is then derived via Eq.~\ref{eq_optlinpol} in a straightforward manner (e.g., \citealt{witzel2011}).

An analysis of \emph{circular} polarization requires the use of a \emph{quarter wave plate} (QWP). The QWP converts circular into linear polarization; the linearly polarized light can be analyzed as discussed above. For a QWP with its axis of minimum index of refraction -- its \emph{fast axis} -- being the $x$ axis in our usual (\S\,\ref{ssect_waves}) coordinate system, its impact on circularly polarized light can be written in Jones calculus like

\begin{equation}
\label{eq_qwp}
\left[\begin{array}{rr} 1 & 0 \\ 0 & i \end{array}\right]\left[\begin{array}{r} 1 \\ \mp i \end{array}\right] = \left[\begin{array}{r} 1 \\ \pm 1 \end{array}\right]
\end{equation}

\noindent
which denotes the application of the Jones matrix of the QWP to a circularly (RHC or LHC) polarized wave, resulting in a linearly polarized wave with diagonal plane of polarization \citep{fowles1975}. Comparison of the result to the definition of the Stokes parameters (\S\,\ref{sssect_stokes}) shows that the QWP converts $V$ to $U$. Accordingly, we can now derive $V$ from an analysis of linear polarization according to Eq.~\ref{eq_optlinpol}, resulting in

\begin{equation}
\label{eq_optcircpol}
\frac{V}{I} = \frac{I(45\deg) - I(135\deg)}{I(45\deg) + I(135\deg)}
\end{equation}

\noindent
(cf., e.g., \citealt{goodrich1995}). I note that the choice of QWP orientation -- here along the $x$ axis -- is arbitrary; for example, a diagonal orientation of the fast axis leads to a conversion from $V$ to $Q$ (e.g., \citealt{fowles1975}).

\subsection{X and $\gamma$ Ray Observations \label{ssect_xray}}

\noindent
Due to the high photon energies and short wavelengths involved, optical elements used at optical wavelengths become transparent at $X/\gamma$ ray energies. Polarimetry at these wavelengths can be based on any of three distinct physical effects.

\begin{figure}[t!]
\centering
\includegraphics[width=82mm]{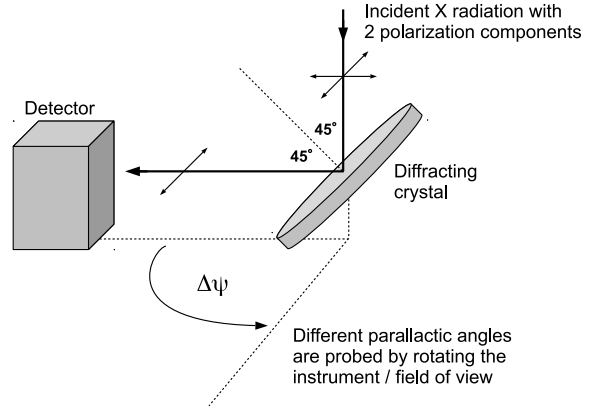}
\caption{Illustration of Bragg diffraction polarimetry.\label{fig_xray-diffraction}}
\end{figure}

\para{Bragg diffraction.} Light with sufficiently short (less than a few nanometers) wavelength $\lambda$ falling onto a crystal is reflected by the crystal according to \emph{Bragg's law}

\begin{equation}
\label{eq_bragg}
n\,\lambda = 2\,d\,\sin\theta ~~; ~~~ n = 1, 2, 3, ...
\end{equation}

\noindent
with $d$ denoting the distance between two consecutive atomic layers measured perpendicular to the surface of the crystal, $\theta$ denoting the angle of incidence measured between the infalling light ray and the surface of the crystal, and $n$ being the order of diffraction \citep{born1999}. Using the relation derived in \S\,\ref{sssect_thomson} it becomes evident that the reflected light is linearly polarized, with the component parallel to the surface prevailing. If incident and reflected rays are perpendicular -- meaning $\theta=45\deg$ -- only the polarization component parallel to the surface of the crystal remains in the reflected light. The linear polarization state of a science target can be derived by rotating the field of view of the instrument and observing the resulting cosinusoidal profile equivalent to Eq.~\ref{eq_rotpol}; I illustrate the measurement geometry in Fig.~\ref{fig_xray-diffraction}. As Bragg's law is strictly valid only for one specific wavelength, this method is, a priori, limited to very narrow energy bands at each order of diffraction. This condition can be relaxed by using ``mosaicked'' crystals composed of many small crystalets, thus providing a range of $d$ values for diffraction \citep{silver2010}.

Bragg diffraction polarimetry was the method used for the only X-ray polarimeter ever implemented in a space telescope, the OSO-8 satellite. It was used to measure the polarization of the Crab nebula \citep{weisskopf1978}, resulting in the ``first and only high-precision X-ray polarization measurement obtained for any cosmic source'' \citep{silver2010}.

\para{Scattering polarimetry.} As discussed in \S\,\ref{sssect_thomson}, Thomson or Compton scattering of photons by electrons is sensitive to the linear polarization of the incident photons. This is exploited in \emph{scattering polarimeters}. We may assume, as usual, partially polarized light composed of photons propagating in $z$ direction. These photons arrive at a \emph{scattering detector} that provides material for scattering the incident light; the scattering detector is located at the origin of the $xy$ plane. A certain fraction of the arriving photons will be scattered at right angles into the $xy$ plane where they are recorded by a \emph{calorimeter}. In case of no polarization, the distribution of scattered photons in the $xy$ plane will be isotropic. If the light is partially linearly polarized, photons will be scattered preferentially perpendicular to the direction of polarization projected onto the $xy$ plane; the resulting distribution is given by expressions equivalent to Eq.~\ref{eq_rotpol} when taking into account the instrument and scattering geometries \citep{mcconnell2010}.

\para{Photoelectron tracking.} Irradiation of X rays on a medium can cause the release of photoelectrons. The direction of photoelectron emission is a function of the polarization of the incident light. Assuming a linearly polarized X ray photon propagating in $z$ direction that causes the emission of a photoelectron at the origin of the $xy$ plane, the differential cross section of photoelectron emission is given by

\begin{equation}
\label{eq_xray-photo}
\frac{{\rm d}\sigma}{{\rm d}\Omega} \propto \frac{\sin^2\theta\,\cos^2\phi}{[1-\beta\cos\theta]^4}
\end{equation}

\noindent
where $\sigma$ is the cross section, $\Omega$ denotes the solid angle, $\beta$ is the electron speed in units of speed of light, $\theta$ is the angle between the path of the incident photon and the path of the emitted photoelectron, and $\phi$ is the angle between the path of the photoelectron and the direction of polarization of the photon projected onto the $xy$ plane \citep{bellazzini2010}. Accordingly, the distribution of photoelectrons is a function of photon polarization, resulting in a characteristic $\cos^2\phi$ pattern in the $xy$ plane.

Compared to classical Thomson scattering, the photoelectric effect is more efficient in analyzing the photon polarization: whereas the differential cross section of Thomson scattering \emph{decreases} with $\theta$ increasing from $0\deg$ to $90\deg$ (cf., e.g., \citealt{rybicki1979}), it strongly \emph{increases} in case of photoelectron emission (Eq.~\ref{eq_xray-photo}). At energies of few keV (i.e., $\beta\lesssim0.1$), the differential cross section peaks at $\theta\approx90\deg$, meaning most of the photoelectrons are emitted within or close to the $xy$ plane. The electron paths can be traced by semiconductor (CCD), gas, or scintillation photo-detectors located in the $xy$ plane; again, the degree and orientation of macroscopic polarization can be derived from the photoelectron distribution via Eq.~\ref{eq_rotpol} (or equivalent expressions).

The use of a certain method for X/$\gamma$ ray polarimetry is largely dictated by the photon energies. Soft X rays can be analyzed with either method; the analysis of hard X and $\gamma$ rays is usually limited to Compton scattering polarimetry (cf., e.g., \citealt{bloser2010}).

\section{SCIENCE CASES \label{sect_science}}  

\subsection{Solar and Stellar Physics \label{ssect_stars}}

\noindent
Across the Hertzsprung-Russell diagram, stars, including the sun, are known to posess magnetic fields with field strengths ranging from a few to several ten thousand Gauss (e.g., \citealt{schrijver2000,berdyugina2009}). In case of hot stars with radiative outer layers (roughly, spectral classes O--A), magnetic fields are supposed to be ``fossil'', i.e., inherited from the intergalactic medium the stars formed from; in case of stars with convective outer layers (approximately spectral classes F--M), magnetic fields are generated by dynamo processes (e.g., \citealt{berdyugina2009}). Accordingly, analyses of stellar magnetic fields are able to constrain solar and stellar dynamo models.

Stellar magnetic fields -- including here also magnetic white dwarfs with $B\lesssim10^9$\,G \citep{jordan2009} -- can be analyzed via spectropolarimetry of absorption lines that are affected by Zeeman splitting (\S\,\ref{ssect_zeeman}). According to Eqs. \ref{eq_zeeman-strong-stokes-m0}, \ref{eq_zeeman-strong-stokes-m+-} and \ref{eq_zeeman-weak-stokes}, the orientation of the field lines can be assessed from the relative strength of linear and circular polarization (see, e.g., \citealt{donati2009} for a review). In case of weak Zeeman splitting (Eq.~\ref{eq_zeeman-weak-stokes}) -- a common case in stellar spectral lines -- the magnetic field strength $B$ enters (via Eq.~\ref{eq_zeeman}) linearly into $V$ and quadratically into $Q$. On the one hand, this makes it possible to estimate $B$ directly from the $V(\nu)$ profile; on the other hand, this complicates the analysis of the field orientation. Spatially resolved maps of magnetic fields of the sun (e.g., \citealt{stenflo2013}) or stars (e.g., \citealt{arzoumanian2011}), usually based on circular polarization, are referred to as \emph{magnetograms}.

Complementary to Zeeman effect measurements, the Hanle effect (\S\,\ref{ssect_hanle}) can be used to probe the magnetic field of the sun \citep{berdyugina2004,milic2012}. This is achieved by simultaneous spectropolarimetric observations of several molecular fluorescent lines; important diagnostic molecules are C$_2$ and MgH. In addition, scattering polarization by Rayleigh and Raman scattering probes the physical conditions in stellar atmospheres (e.g., \citealt{sampoorna2013}).

\subsection{Planetary System Bodies \label{ssect_planets}}

\subsubsection{Solid Surfaces}

\noindent
Sunlight reflected at a solid surface -- like the ones of rocky planets or asteroids -- becomes partially linearly polarized due to scattering polarization (\S\,\ref{ssect_scattering}). Unsurprisingly, the observed degree of polarization is a function of the relative position of observer, reflecting body, and the star (the \emph{phase angle} in case of the solar system). The \emph{maximum} degree of linear polarization, in the following denoted with $\bar m_L$, is a function of wavelength and of the structure of the reflecting material. The interplay between absorption and scattering of light causes the \emph{Umov effect}, a characteristic anti-correlation between $\bar m_L$ and geometric albedo $\mathcal{A}$ of a solid surface \citep{bowell1974}. For a given material -- like lunar regolith, sand, basalt, or granite -- polarization and albedo are related like

\begin{equation}
\label{eq_umov}
\log({\mathcal{A}}) = -c_1 \log(\bar m_L) + c_2
\end{equation}

\noindent
at visible wavelengths, with constants $c_1\approx1$ and $c_2\approx-2$ (for ${\mathcal{A}}, m_L \in[0.01,1]$). Notably, this implies degrees of polarization close to 100\% for very low albedos. Observed deviations from this relation indicate a change in the structure of the surface material; accordingly, ${\mathcal{A}}-\bar m_L$ diagrams can be used to assess the surface composition of a planet or any other solid body. In addition, characteristic variations of $\bar m_L$ with time can be used to estimate the rotation period and/or surface profile of a small body (e.g., an asteroid) that is not resolved spatially by observations (e.g., \citealt{ishiguro1997,cellino2005}).

Radar astronomical observations \citep{ostro1993,campbell2002} exploit the polarization state of the reflected radio light. The transmitted radar signal has a well defined polarization state (usually 100\% circular). In case of a single reflection at an ideal dielectric surface, the circular polarization state of the echo signal is inverted with respect to the transmitted signal, the linear polarization state (expressed via Stokes $Q$ by proper choice of coordinates) remains unchanged (cf. Eq.~\ref{eq_mueller-matrix-reflect}). Multiple scattering and/or refraction at rough surfaces lead to some of the echo light being in the same circular polarization state and/or inverted linear polarization state compared to the infalling light. Denoting the polarization states as the ``same'' (S) and ``opposite'' (O) ones with respect to the transmitted radar signal, one can define the polarization ratios

\begin{equation}
\label{eq_radar}
{\mathcal{R}}_{C} = \frac{\Sigma_{\rm SC}}{\Sigma_{\rm OC}} ~~~~~ {\rm and} ~~~~~ {\mathcal{R}}_{L} = \frac{\Sigma_{\rm OL}}{\Sigma_{\rm SL}}
\end{equation}

\noindent
with ``L'' and ``C'' referring to linear and circular polarization, respectively, and $\Sigma$ denoting the radar cross section of the target. Accordingly, both ${\mathcal{R}}_{L}$ and ${\mathcal{R}}_{C}$ would be zero for an ideal smooth surface. For most solar system objects, ${\mathcal{R}}_{C}\lesssim0.3$, with the notable exception of the icy moons of Jupiter for which ${\mathcal{R}}_{C}\gtrsim1$ \citep{ostro1993}.

\subsubsection{Atmospheres}

\noindent
Reflection of light at (sufficiently dense) planetary atmospheres (e.g., \citealt{buenzli2009}, and references therein) is mainly affected by two (linearly) polarizing processes: (i) Rayleigh scattering at molecules and aerosol haze particles, and (ii) refraction and reflection at liquid droplets in clouds. Whereas individual interactions can lead to degrees of linear polarization up to 100\%, the signal observed by a distant observer is the average over multiple light rays, which partially averages out the polarization signal and reduces the observed degree of polarization. The actual polarization levels depend strongly on the reflection geometry and the chemical composition of the atmosphere. Within and around the regime of visible wavelengths, observed levels of polarization -- integrated over the planetary disks -- are $<$5\% for Venus, 5--10\% for Jupiter and Saturn, and up to $\approx$50\% for Titan (Saturn's moon). 

Quantitative investigations of the polarization properties of planetary atmospheres require numerical modeling (e.g., \citealt{buenzli2009}). The polarization of -- intrinsically unpolarized -- starlight reflected from planets is used for direct imaging of exoplanets via polarimetric differential imaging (e.g., \citealt{milli2013}).

\subsection{Interstellar Matter \label{ssect_ism}}

\noindent
Interstellar space is filled with diffuse matter occurring in a large variety of states, from cold dense molecular (main species being H$_2$, CO, and H$_2$O) clouds with temperatures $T$ of few Kelvin and (hydrogen) particle densities $10^{3...5}$\,cm$^{-3}$ up to the hot ionized (coronal) medium (main species being \myion{H}{ii}, \myion{C}{iv}, \myion{N}{v}, and \myion{O}{vi}) with $T\approx10^6$\,K and hydrogen densities $\approx3\times10^{-3}$\,cm$^{-3}$. In addition, interstellar dust is omnipresent throughout galaxies (see, e.g., \citealt{kwok2007} for a detailed overview).

The interplay of interstellar dust and galactic magnetic fields (\S\S\,\ref{sssect_dust}, \ref{ssect_galfields}) is responsible for the \emph{interstellar polarization} of scattered starlight (see, e.g., \citealt{das2010,matsumura2011} for recent discussions); the degree of linear polarization is approximately given by Serkowski's law and, accordingly, ranges from a few to about ten per cent \citep{draine2003}. In case of \emph{circumstellar} material in the immediate vicinity of a star, scattering polarization can arise from:

\begin{itemize}

\item[(i)]  The alignment of dust grains in the magnetic field of a circumstellar disk or star-forming nebula.

\item[(ii)]  Scattering at spherical or randomly oriented dust grains; in this case, polarization arises from geometry because the incident light arrives from a well-defined direction -- the star.

\item[(iii)]  Scattering at magnetically aligned dust grains plus dichroic absorption by foreground material, leading (also) to \emph{circular} polarization with $m_C\lesssim20$\% \citep{kwon2013}.

\end{itemize}
In case (i), infrared polarimetric imaging has revealed the magnetic field structures in disks around young stars as well as characteristic ``hour-glass'' field geometries in star forming regions (e.g., \citealt{cho2007,sugitani2010}). In case (ii), polarimetric imaging of circumstellar material shows a highly symmetric circular pattern centered at the star, with the orientation of polarization being perpendicular to the direction of the incident radiation. This has been used to analyze (proto)stars embedded in dense interstellar matter (e.g., \citealt{saito2009}). Circular scattering/absorption polarization (case iii) has been observed only in a few star forming regions \citep{kwon2013}.

\subsection{Astrobiology \label{ssect_bio}}

\noindent
Complex helical organic molecules in terrestrial life forms -- like amino acids -- show \emph{homochirality}: out of two helix orientations possible, only one is used exclusively. This phenomenon implies that one of the two orientations was preferred in pre-biotic chemistry. A possible cause is circularly polarized light in star forming regions (\S\,\ref{ssect_ism}) leading to preferential photo-dissociation of organic molecules with one specific orientation. This causes an excess of molecules with a given orientation and, eventually, on Earth to which organic matter is transported via comets and meteoroids \citep{demarcellus2011,kwon2013}.

Homochirality causes light reflected from certain biological surfaces to be circularly polarized (cf. \S\,\ref{ssect_chirality}). This effect can -- in principle -- be exploited for detecting life on other planets via (spectro)polarimetry of starlight reflected from the surface (e.g., \citealt{sparks2012}).

\subsection{Astronomical Masers \label{ssect_masers}}

\noindent
Stimulated emission of radiation at radio frequencies -- maser radiation -- can be observed from the interstellar matter in star-forming regions and from the circumstellar envelopes of late-type (super)giant stars (e.g., \citealt{kwon2012}). Maser radiation is emitted as molecular line emission with very high brightness temperatures up to roughly $10^{12}$\,K. Species known to act as astrophysical maser media are the molecules OH, H$_2$O, CH$_3$OH, NH$_3$, HC$_3$N, H$_2$CO, CH, SiO, SiS, and HCN, plus atomic hydrogen (H). Maser radiation has been observed at frequencies from 1.61\,GHz (from OH) to 662.4\,GHz (from H), i.e., across the entire radio regime \citep{reid1981,elitzur1982,townes1997}.

Astronomical masers tend to show substantial Zeeman line splitting caused by magnetic fields permeating the maser medium. In the case of strong ($\Delta\nu_{\rm z}\gg\Delta\nu$) Zeeman splitting, the resulting line polarizations are given by Eqs.~\ref{eq_zeeman-strong-stokes-m0} and \ref{eq_zeeman-strong-stokes-m+-}. In case of weak Zeeman splitting ($\Delta\nu_{\rm z}\ll\Delta\nu$) however we have to take into account that maser radiation is caused by \emph{stimulated, coherent} emission, meaning a coherent superposition of electric waves. For a \emph{single electromagnetic wave}, the resulting Stokes parameters, in units of Stokes $I$, are

\begin{eqnarray}
\label{eq_maser-weak-stokes}
\frac{Q}{I} & = & -1 + \frac{2}{3\sin^2\theta}  \\
\frac{U}{I} & = & \pm\frac{2}{3\sin^2\theta}\left(3\sin^2\theta - 1\right)^{1/2}  \nonumber \\
\frac{V}{I} & = & 0 \nonumber
\end{eqnarray}

\noindent
with $\theta$ denoting the angle between the magnetic field and the line of sight \citep{elitzur1991,elitzur2000}. For $\sin^2\theta\leq1/3$ (i.e., $\theta\lesssim35\deg$), $Q/I=1$ and $U/I=0$. As $(Q/I)^2 + (U/I)^2 = 1$ for all $\theta$, a wave emitted by an ideal maser is always fully linearly polarized. When averaging over \emph{multiple waves} -- as in any realistic astronomical observation -- the sign ambiguity in $U/I$ causes Stokes $U$ to average out; only $Q$ remains, implying a partial linear polarization with $m_L=Q/I$. However, more recent calculations based on numerical simulations of realistic maser radiation fields find that the analytical estimates quoted above suffer from over-simplifications; the actual levels of linear polarization should be substantially smaller than the ones predicted by Eq.~\ref{eq_maser-weak-stokes} \citep{dihn2009}. Furthermore, already moderate ($\Delta\nu_z<\Delta\nu$) Zeeman splitting introduces \emph{circular} polarization with amplitudes as high as $m_C\approx20$\%, with frequency-dependent profiles $V(\nu)$ similar to Eq.~\ref{eq_zeeman-weak-stokes} for sufficiently small $\theta\lesssim30\deg$ \citep{elitzur2000,dihn2009}.

\subsection{Pulsars \label{ssect_pulsars}}

\noindent
Pulsars are neutron stars with strong magnetospheres. Their radiation is composed of thermal radiation from the neutron star surface -- at temperatures $T\approx10^6$\,K -- and, predominantly, non-thermal synchrotron and curvature radiation created within the stellar magnetosphere. The observational pulsar phenomenology is given by geometry: the magnetic axis of the star is tilted relative to its spin axis. If the magnetic axis points to the observer during a rotation period, the radiation from the magnetosphere becomes visible as a short pulse of light. Observed pulse periods are located roughly in the range from few milliseconds to tens of seconds, with most pulsars having periods about few hundred milliseconds. To date, approximately 2000 pulsars are known which are distributed throughout the Milky Way (see, e.g., \citealt{lyne2012} for a review).

The magnetic field of the neutron star can be assumed to be a relic of the field of the progenitor star. Conservation of magnetic flux demands very high field strengths nearby the star, with values in the range $B\approx10^{6-10}$\,T. The field geometry is bipolar (at least within the \emph{light cylinder}, i.e. the regime of co-rotation speeds below the speed of light). The combination of strong magnetic field plus fast rotation leads to the creation of a strong electric field at the stellar surface, with field strengths up to $E\approx10^{12}$\,V\,m$^{-1}$. The electric field extracts charged particle (electrons, ion) at and around the magnetic poles. The charges propagate along the magnetic field lines at highly relativistic (Lorentz factors $\gamma\approx10^7$) energies. Those primary electric charges, plus secondary charges with $\gamma\approx1000$ originating from electron-positron pair creation, produce synchrotron and (mostly) curvature radiation directed along the magnetic field lines. The emission geometry provides the ``lighthouse effect'' necessary for the observational pulsar phenomenology. The emitted radiation is partially coherent; the highest flux densities are usually observed at low -- few GHz -- radio frequencies \citep{michel1991,beskin1993}.

By geometry, the radiation from pulsars can roughly be approximated as synchrotron radiation from collimated beams of electrons (\S\,\ref{ssect_synchrotron}). Accordingly, one observes (e.g., \citealt{rankin1983}) both linear and circular polarization approximately following the pattern outlined in Fig.~\ref{fig_synchpol}, with details depending on the actual viewing geometry. The angle of linear polarization ``swings'' through a range of values during a pulse because of the rotation of the star \citep{michel1991}. Historically, polarimetric observations of the Crab nebula, the supernova remnant surrounding the Crab pulsar, provided the first evidence ever for synchrotron radiation from astronomical objects \citep{oort1956}.

\subsection{Active Galactic Nuclei \label{ssect_agn}}

\noindent
With luminosities up to approximately $10^{15}\,L_{\odot}$, active galactic nuclei (AGN; see, e.g., \citealt{beckmann2012} for a recent review) are the most luminous persistent objects in the universe. Their source of energy is the accretion of interstellar matter onto supermassive -- meaning $M_{\bullet}\approx10^{6-10}\,M_{\odot}$ -- black holes which are located in the centers of most, if not all, galaxies. The energy gained from accretion is (largely) radiated away in the form of broad-band continuum emission that is observed from low-frequency radio to high-energy $\gamma$ energies. AGN emission shows strong variability and characteristic statistical properties (e.g., \citealt{park2012,kim2013}). The radiation from AGN crudely falls into two physical regimes. At low energies ranging roughly from radio to ultraviolet frequencies, the emission is dominated by synchrotron radiation. At higher energies, the radiation is probably produced by inverse Compton scattering of low-energy synchrotron photons.

As AGN are synchrotron sources, their emission is linearly polarized; see also the example provided by Fig.~\ref{fig_optpol}. Accordingly, AGN polarization has been studied extensively for several decades and has been used to address the geometries of magnetic fields and the matter distributions (notably particle densities via Faraday rotation) in and around active nuclei (see, e.g., \citealt{saikia1988} for an overview). Degrees of \emph{linear} polarization are $m_L\lesssim20$\%, with typical values around $m_L\approx5$\% (e.g., \citealt{trippe2010,trippe2012a}). \emph{Circular} polarization has been observed in a handful of sources on levels $m_C\lesssim1$\% (e.g., \citealt{agudo2010}).

The outflows of matter from AGN, especially the formation of collimated jets which extend over several megaparsecs in extreme cases, are intimately linked to the immediate (tens of Schwarzschild radii) environment of the central black hole and the geometry of the magnetic fields located there (e.g., \citealt{narayan2005}). AGN jets are -- largely -- optically thin emitters of synchrotron radiation best observable at radio frequencies. Accordingly, linear polarization is used to trace the orientation and strength of magnetic fields along the jets.\footnote{At this point it is important to note that the observed polarization of extended sources is a function of angular resolution: if several individual emitters of polarized radiation fall within the same resolution element (the \emph{point spread function} or \emph{beam} of the instrument), the polarization signal can be averaged out partially -- a phenomenon known as \emph{beam depolarization}.} Observations at multiple wavelengths permit the use of Faraday rotation and Faraday depolarization as a probe of magnetic fields and matter distributions (e.g., \citealt{macquart2006,taylor2006,trippe2012b}). A somewhat unexpected property of AGN jets was the discovery of \emph{inverse depolarization} -- higher degrees of linear polarization at longer wavelengths -- in some sources which has been interpreted as a ``conspiracy'' of spatial small-scale structure and Faraday rotation (e.g., \citealt{homan2012}). Occasional observations of circular polarization in AGN jets, most notably in 3C\,84 with polarization levels up to $m_C\approx3$\%, have been attributed to polarization conversion \citep{homan2004}.

Historically, polarimetric observations of the active Seyfert galaxy NGC~1068 helped to establish the -- nowadays standard -- \emph{viewing angle unification scheme} of AGN. Spectropolarimetry at optical wavelengths shows that the total flux received from the galaxy is actually composed of two components: one -- unpolarized -- from directly observable gas with narrow emission lines, one -- linearly polarized -- from gas with much broader emission lines located within a dust torus and visible only indirectly via Thomson scattering toward the observer (cf., e.g., \citealt{baek2007,lee2011}). This observation eventually removed the distinction between narrow and broad emission line galaxies which were found to be different realizations of AGN \citep{miller1983,miller1991}.

\subsection{Galactic Magnetic Fields \label{ssect_galfields}}

\noindent
Disk galaxies and clusters of galaxies are permeated by large-scale (many kpc) magnetic fields with field strengths $B$ on the order of $\mu$Gauss. In case of disk galaxies, these fields are aligned with the galactic plane and follow closely the galactic structure, especially spiral arms (see, e.g., \citealt{fletcher2011} for an impressive example). The fields are supposed to be generated via amplification of primordial cosmic magnetic fields -- with $B\sim10^{-20}$\,G -- by ``galactic dynamos'' driven by galactic rotation. The most widely applied model is the \emph{$\alpha$--$\Omega$ disk dynamo} which comprises as parameters (i) the angular speed $\Omega$ of galactic rotation and (ii) the quantity $\alpha=-\tau({\bf v}\cdot\nabla\times{\bf v})/3$, with $\tau$ being the decorrelation time of plasma turbulences and {\bf v} being the plasma velocity (the expression in brackets is also referred to as ``kinematic helicity''). In case of galaxy clusters, the fields supposedly originate from extended AGN jets (\S\,\ref{ssect_agn}) which carry strong magnetic fields into the intragalactic medium and where these are dissolved over time \citep{wielebinski1993,kulsrud2008}.

The analysis of large-scale magnetic fields is based on signatures of their interaction with the interstellar or intergalactic medium, specifically:

\begin{itemize}

\item[(i)]  Faraday rotation (\S\,\ref{sssect_faraday-rot}) of radiation from pulsars or extragalactic background sources (cf. \citealt{clarke2004,kronberg2004,kronberg2011}).

\item[(ii)]  Weak Zeeman effect line splitting, especially the $V(\nu)$ profiles in \myion{H}{i} emission and absorption lines (\S\,\ref{ssect_zeeman}; cf. \citealt{heiles2009}).

\item[(iii)]  Polarized synchrotron radiation (\S\,\ref{ssect_synchrotron}; cf. \citealt{heald2009}).

\item[(iv)]  Polarization arising from scattering at magnetically aligned dust grains (\S\,\ref{sssect_dust}; cf. \citealt{pavel2011}). Notably, this method provided detailed insight into the magnetic field geometry within the center of the Milky Way \citep{nishiyama2010} and in parts of the Large Magellanic Cloud \citep{jkim2011}.

\item[(v)]  The Chandrasekhar-Fermi effect (\S\,\ref{sssect_chandra}; \citealt{chandra1953})

\end{itemize}

\noindent
As should be clear from the discussion provided in \S\S\,\ref{sect_polsources} and \ref{ssect_matter}, (i) and (ii) provide information on magnetic field components along the line of sight, whereas (iii), (iv), and (v) provide information on field components perpendicular to the line of sight.

\subsection{Gamma Ray Bursts \label{ssect_grb}}

\noindent
Gamma-ray bursts (GRB; e.g., \citealt{piran2005,gehrels2009}) are short, intense pulses of soft (hundreds of keV) $\gamma$ rays of cosmological origin occurring a few times per day. GRBs last from fractions of a second to hundreds of seconds; with luminosities up to about $10^{46}$\,W they are among the most luminous (transient) sources of radiation in the universe. According to their duration, GRBs fall into either of two groups:

\epara{Long GRBs} typically last tens of seconds and are associated with type Ib/c supernovae. They are assumed to originate from \emph{collapsars}, massive evolved stars (probably Wolf-Rayet stars) whose cores collapse into black holes.

\epara{Short GRBs} usually last less than one second. They are assumed to be caused by mergers of compact objects in binary systems, like two neutron stars or one neutron star and one stellar black hole.

In either case, the outflowing plasma is collimated into relativistic jets with opening angles of a few degrees; this explains the very large apparent isotropic luminosities of GRBs. The GRB emission results from synchrotron and inverse Compton radiation from relativistic electrons. The combination of synchrotron radiation (\S\,\ref{ssect_synchrotron}) and non-isotropic geometry should lead to substantial linear polarization, and, indeed, degrees of polarization up to 30\% have been reported \citep{goetz2013,mundell2013}. Sufficient measurement accuracies provided, the polarization can be used to probe the plasma-physical conditions and magnetic fields in GRBs similar to the procedures for AGN (\S\,\ref{ssect_agn}).

As noted by, e.g., \citet{toma2012}, polarized high-energy emission from cosmological sources like GRBs can be used to probe a vacuum birefringence arising from a violation of the Lorentz invariance of Einstein's theory of relativity.

\subsection{Cosmic Background Radiation \label{ssect_cmb}}

\noindent
The cosmic microwave background (CMB) is supposed to originate from the hot plasma filling the universe approximately 400\,000 years after the big bang. To first order, the CMB corresponds to thermal emission from a black body with a temperature of $\approx$2.7\,K. Plasma density fluctuations imprint characteristic fluctuations with amplitudes on scales of $\mu$K into the angular distribution of the CMB. In addition to fluctuations in the total intensity, one may expect localized linear scattering polarization \emph{if} the radiation propagating through the plasma shows quadrupole anisotropies -- differences in intensities at angles of $90\deg$ in the sky plane. Depending on the underlying geometry, two signatures or \emph{modes} of polarization have to be distinguished \citep{zaldarriaga1997,kamionkowski1997}.

\para{E mode polarization.}  A polarization geometry where the orientations of polarization are perpendicular to the gradient of a local perturbation of the CMB is referred to as \emph{electric-field like} (hence E) or \emph{gradient mode} (G) polarization. By construction, such a polarization pattern does not show a handedness. E mode polarization can be attributed to local energy density fluctuations, also known as \emph{scalar} perturbations.

\para{B mode polarization.} A local curl pattern of polarization with distinct handedness is referred to as \emph{magnetic-field like} (hence B) or \emph{curl mode} (C) polarization. The amplitudes of those patterns are supposed to be roughly one order of magnitude weaker than those of E mode signatures. By geometry, B mode polarization requires \emph{tensor} perturbations of the CMB. Those perturbations occur due to the propagation of gravitational waves through the CMB plasma; accordingly, measurements of B mode polarization are a key experiment for probing primordial gravitational waves and cosmic inflation theories.

In the past decade, E mode polarization has been observed by a variety of ground based CMB telescopes in the approximate frequency range 30--150\,GHz (e.g., \citealt{leitch2002,kovac2002,park2002,readhead2004,takahashi2010}). A typical CMB telescope is designed as an interferometer with multiple receivers located on a common carrier platform spanning a few meters in diameter. By design (using Rayleigh's criterion for angular resolution) CMB telescopes are sensitive to structures on angular scales of about 1--2$\deg$, i.e. the characteristic size scale of E mode polarization patterns. More recently, the \name{Planck} satellite has begun a polarization monitoring program aimed at both E and B modes (see, e.g., \citealt{lamarre2003} for technical details), and an observation of B mode polarization by a ground-based CMB telescope has been reported \citep{hanson2013}.

\section{CONCLUSIONS \label{sect_conclude}}  

\noindent
Polarization of light and polarimetry play fundamental roles in astrophysics. Polarization is fundamentally linked to the internal \emph{geometry} of sources of radiation: the strengths and orientations of magnetic fields, the distribution and orientation of scattering particles like dust grains, the microscopic structure of reflecting surfaces, or intrinsic anisotropies of the primordial plasma filling the early universe. Accordingly, polarimetry has found application in a vast variety of astrophysical fields of study ranging all the way from solar physics to cosmology, comprising even a ``personal touch'': Understanding the interplay between circularly polarized starlight and the interstellar medium might help to understand the formation of life on Earth.

Reviewing its applications, it is evident that polarimetry is a powerful tool for astrophysics; it provides rich information on the physics of targets that cannot be obtained in any other way. Consequently, a large number of dedicated observational instruments has been constructed and progress is fast. One important current trend is the development of instruments dedicated to polarimetry at the high-energy end of the electromagnetic spectrum, at X and $\gamma$ ray wavelengths (\S\,\ref{ssect_xray}); another one could be investigations of optical polarimetric interferometry \citep{elias2001}. Each new technical development eventually opens new windows for observational astronomy. Likewise, more traditional polarimetric techniques profit from the general progress in instrumentation technologies; a key aspect is the improvement of instrumental sensitivities -- which have always been harmed by polarimetry measuring relatively small differential fluxes by definition. This being said, we may conclude that polarimetry has the potential for new and exciting astrophysical discoveries in the future.

\addcontentsline{toc}{section}{Acknowledgments}
\acknowledgments{\noindent I am grateful to \name{Myungshin Im}, \name{Masateru Ishiguro}, \name{Junghwan Oh}, \name{Taeseok Lee}, \name{Jong-Ho Park}, and \name{Jae-Young Kim} (all at SNU) for valuable discussion. I acknowledge financial support from the Korean National Research Foundation (NRF) via Basic Research Grant 2012-R1A1A2041387. Last but not least, I am grateful to an anonymous referee for valuable comments.}

\end{document}